\documentclass[11pt]{article}

\usepackage{graphicx}
\usepackage{amsmath}
\usepackage{amssymb}
\usepackage[english,francais]{babel}

\newfont{\ensmathquatorze}{msbm10 scaled 1400}
\newfont{\ensmathonze}{msbm10 scaled 1100}
\newfont{\ensmathdix}{msbm10}
\newfont{\ensmathneuf}{msbm10 scaled 833}
\newfont{\ensmathhuit}{msbm10 scaled 694}
\newfam\ensmathfam                        
\textfont\ensmathfam=\ensmathonze        
\scriptfont\ensmathfam=\ensmathdix       
\scriptscriptfont\ensmathfam=\ensmathhuit
\def\ensmf{\fam\ensmathfam\ensmathonze}         
\def\be{\begin{equation}}
\def\ee{\end{equation}}

\def\bea{\begin{eqnarray}}
\def\eea{\end{eqnarray}}
\def\beann{\begin{eqnarray*}}
\def\eeann{\end{eqnarray*}}

\def\typ{\mbox{\tiny typ}}

\renewcommand{\leq}{\leqslant}
\renewcommand{\geq}{\geqslant}

\def\eqlaw{\stackrel{\mbox{\tiny\rm (law)}}{=}}     
\newcommand{\ket}[1]{|\kern.3ex#1\kern.3ex\rangle}
\newcommand{\bra}[1]{\langle\kern.3ex #1 \kern.3ex|}
\newcommand{\APPROX}[1]{                
   {{\raisebox{-.3cm}{$\textstyle\simeq$}} \atop {\scriptstyle{#1}}}}

\newcommand{\leadto}[1]{                
   {{\raisebox{-.3cm}{$\textstyle\longrightarrow$}} \atop {\scriptstyle{#1}}}}
\newcommand{\mean}[1]{\left\langle #1 \right\rangle} 
\newcommand{\smean}[1]{\langle #1 \rangle} 

\newcommand{\EXP}[1]{{\mbox{\large e}}^{#1}}         

\newcommand{\im}{\mathop{\mathrm{Im}}\nolimits}      
\newcommand{\tr}[1]{\mathop{\mathrm{Tr}}\nolimits\left\{ #1 \right\}}  

\newcommand{\proba}{\mathop{\mathrm{Prob}}\nolimits}  
\newcommand{\heav}{\mathop{\mathrm{\rm Y}}\nolimits}  

\def\RR{{\ensmf R}}                 

\def\I{{\rm i}}                  
\def\D{{\rm d}}                  
\def\Dc{{\rm D}}                 

\newcommand{\drond}[2]{\frac{\partial #1}{\partial #2}} 

\newcommand\ab{{\alpha\beta}}
\newcommand\ba{{\beta\alpha}}

\newcommand\lab{l_{\alpha\beta}}

\newcommand\ide{\mbox{\bf 1}}

\newcommand{\diagram}[3]{\raisebox{#3}{\includegraphics[scale=#2]{#1}}}

\def\sw{S_{\rm w}}

\newcommand{\cab}{\chi_{\alpha\beta}}
\newcommand{\cba}{\chi_{\beta\alpha}}

\newcommand{\sg}{\sqrt {\gamma }}

\newcommand{\wal}{W_{\alpha\beta}}

\begin{document}
\selectlanguage{english}


\title{Functionals of the Brownian motion,\\ localization and metric graphs}

\author{Alain Comtet$^{(a,b)}$, Jean Desbois$^{(a)}$ and 
        Christophe Texier$^{(a,c)}$}

\date{July 22, 2005}

\maketitle

\noindent
\hspace{1cm}
\begin{minipage}[t]{13cm}
{\small

  \noindent
  $^{(a)}$
  \begin{minipage}[t]{12.5cm}
    Laboratoire de Physique Th\'eorique et Mod\`eles Statistiques,
    UMR 8626 du CNRS.\\
    Universit\'e Paris-Sud,
    B\^at. 100, F-91405 Orsay Cedex, France.
  \end{minipage}

  \vspace{0.15cm}

  \noindent
  $^{(b)}$
  \begin{minipage}[t]{12.5cm}
    Institut Henri Poincar\'e, 11 rue Pierre et Marie Curie, F-75005 Paris,
    France.
  \end{minipage}

  \vspace{0.15cm}

  \noindent
  $^{(c)}$
  \begin{minipage}[t]{12.5cm}
    Laboratoire de Physique des Solides,
    UMR 8502 du CNRS.\\
    Universit\'e Paris-Sud,
    B\^at. 510, F-91405 Orsay Cedex, France.
  \end{minipage}
  }
\end{minipage}

\begin{abstract}
We review several results related to the problem of a quantum
particle in a random environment.

In an introductory part, we recall how several functionals of the Brownian
motion arise in the study of electronic transport in weakly disordered
metals (weak localization).

Two aspects of the physics of the one-dimensional strong localization are
reviewed~:  some properties of the scattering by a random potential (time
delay distribution) and a study of the spectrum of a random potential on a
bounded domain (the extreme value statistics of the eigenvalues).

Then we mention several results concerning the diffusion on graphs, and more
generally the spectral properties of the Schr\"odinger operator on graphs. The
interest of spectral determinants as generating functions characterizing the
diffusion on graphs is illustrated.

Finally, we consider a two-dimensional model of a charged particle coupled to
the random magnetic field due to magnetic vortices. We recall the connection
between spectral properties of this model and winding functionals of the
planar Brownian motion.
\end{abstract}

\noindent
PACS numbers~: 72.15.Rn~; 73.20.Fz~; 02.50.-r~; 05.40.Jc.





\tableofcontents


\section{Introduction}

\subsection{Weak and strong localization}

At low temperature, the electric conductivity $\sigma$ of metals and weakly
disordered semiconductors is determined by the scattering of electrons on
impurities. It is given by the Drude formula
\be
\sigma_0=\frac{n_e e^2 \tau_e}{m},
\ee
where $e$ and $m$ are the charge and mass of the electron respectively. $n_e$
is the electronic density and $\tau_e$ the elastic scattering time\footnote{
  The elastic scattering time $\tau_e$ is the time characterizing the 
  relaxation of the direction of the momentum of the electron. The 
  conductivity is proportional to $\tau_e$ for isotropic scattering by 
  impurities. For anisotropic scattering the conductivity involves a 
  different time  $\tau_{\rm tr}$ called the ``transport'' time
  \cite{AshMer76}
  (see ref. \cite{AkkMon04} for a discussion within the perturbative approach).
}. 
This purely classical formula is only valid in a regime where quantum
mechanical effects can be neglected. This is the case if the elastic mean free
path of electrons $\ell_e=v_F\tau_e$ is large compared with the De Broglie
wave length $\lambda_F= 2\pi\hbar /mv_F$ corresponding to  the
Fermi energy ($v_F$ is the Fermi velocity).

\vspace{0.25cm}

\noindent
{\it Strong localization.--} When these two length scales
are of the same order $\ell_e\approx\lambda_F$ (strong disorder) the fact that
the electrons are quantum objects must be taken into account and the wave like
character of these particles is of primary importance. It is indeed this wave
character which is responsible for the localization phenomenon. The multiple
scattering on impurities distributed randomly in space creates random phases
between these different waves which can interfere destructively. These
interference effects reduce the electronic conductivity. In the extreme case
of very strong disorder the waves do not propagate anymore and the system
becomes insulating. This is the {\it strong localization} phenomenon which was
conjectured by Anderson in 1958.

\vspace{0.25cm}

\noindent
{\it Weak localization.--} In the 80's it was realized that
even far from the strong localization regime the quantum transport is
affected by the disorder. Diagrammatic techniques used in the weak 
disorder limit  $\lambda_F\ll\ell_e$, were initiated by
the works of Al'tshuler, Aronov, Gor'kov, Khmel'nitzki{\u\i}, Larkin and Lee
\cite{GorLarKhm79,AltKhmLarLee80} (see ref.~\cite{AltLee88} for an 
introduction and ref.~\cite{AkkMon04} for a recent presentation). 
In this regime, called the {\it weak localization} regime, the
Drude conductivity gets a small sample dependent correction whose
average, denoted $\smean{\Delta\sigma}$, is called the ``weak localization
correction''. $\smean{\cdots}$ denotes averaging with respect to the 
random potential.
From the experimental side, this phenomenon is well 
established and has been the subject of many studies (see 
refs.~\cite{Ber84,AroSha87} for review articles).

\vspace{0.25cm}

\noindent
{\it Phase coherence and dimensional reduction.--} The localization phenomenon
comes from the interplay between the wave nature of electronic transport and
the disorder. This manifestation of quantum interferences requires that the
phase of the electronic wave is well defined, however several mechanisms limit
the phase coherence of electrons in metals, among which are the effect of the
vibrations of the crystal (electron-phonon interaction) or the
electron-electron interaction. We introduce a length scale $L_\varphi$, the
phase coherence length, that characterizes the length over which phase
breaking phenomenon becomes effective. The lack of phase coherence in real
systems is the reason why the strong localization regime has not been
observed in experiments on metals. In dimension $d=3$ the strong localization
regime is only expected to occur for sufficiently strong
disorder\footnote{
  Following the scaling ideas initiated by Thouless \cite{LicTho75} and
  Wegner \cite{Weg76,Weg79} it was shown in ref.~\cite{AbrAndLicRam79} that 
  the localization-delocalization transition exists only for dimension 
  $d\geq3$.
  In $d=1$ and $d=2$, the fully coherent system is always strongly localized,
  whatever the strength of the 
  disorder is.\label{Wegner}
}. 
However, strong localization can also be observed in the weak disorder
limit ($\lambda_F\ll\ell_e$) by reducing the dimensionnality\footnote{
  The effective dimension of the system is obtained by comparing the sample 
  size with the phase coherence length $L_\varphi$. For example, a long 
  wire of length $L$ and of transverse dimension $W$ is effectively in 
  a 1d regime if $W\ll L_\varphi\ll L$.
}. It has been shown in the framework of random matrix
theory that the localization length of a weakly disordered quasi-1d wire
behaves as $\lambda\sim N_c\ell_e$, where $N_c$ is the number of
conducting channels\footnote{
  The localization length predicted by the random matrix theory (RMT) can 
  be found in ref.~\cite{Bee97}~: 
  $\lambda_{\rm RMT}=[\beta(N_c-1)+2]\ell_e$, 
  where $\beta=1,2,4$ is the Dyson index describing orthogonal, unitary
  and symplectic ensembles, respectively. 
  Note that one must be careful with the coefficient involved in 
  this relation since the definitions of $\lambda_{\rm RMT}$ and $\ell_e$
  differ slightly in RMT and in perturbation theory.
} \cite{Dor82,Dor88,MelPerKum88,Bee97}.
Therefore the strong localization regime is expected to occur when coherence is
kept at least over a scale $\lambda$. At low temperature, in the absence of
magnetic impurity, phase breaking mechanisms are dominated by
electron-electron interaction, which leads to a divergence of the phase
coherence length $L_\varphi(T)\propto(N_c/T)^{1/3}$ predicted in
ref.~\cite{AltAroKhm82} and verified in several experiments like
\cite{WinRooChaPro86,ThoPepAhmAndDav86,EchGerBozBogNil93b}
(see ref.~\cite{PieGouAntPotEstBir03} for recent measurements down to 
$40\:$mK). 
Therefore the
temperature below which strong localization might be observable is given by
$L_\varphi(T_*)\sim\lambda$, which leads to $T_*\sim1/(N_c^2d\,\tau_e)$.
The crossover temperature in metals is out of the experimental range, however
it becomes reachable in wires etched at the interface of two semiconducting
materials, when the number of conducting channels is highly reduced, and the
manifestation of the strong localization has been observed in 
ref.~\cite{GerKhaMikBozBog97}.

\vspace{0.25cm}

\noindent
{\it Strictly one-dimensional case.--} In a weakly disordered and
coherent quasi-1d wire, the weak localization regime only occurs for length
scales intermediate between the elastic mean free path and the localization
length $\ell_e\ll L\ll\lambda$. In the strictly one dimensional case,
since $\lambda\simeq4\ell_e$ for weak disorder\footnote{
  The elastic mean free path $\ell_e=v_F\tau_e$ is given by the self 
  energy $1/(2\tau_e)=-\im\Sigma^{\rm R}(E)$. For example, for a Gaussian
  disorder with local correlations, $\smean{V(x)V(x')}={w}\delta(x-x')$, 
  we obtain $1/\tau_e\simeq2\pi\rho_0{w}$ for a weak disorder, where 
  $\rho_0$ is the free density of states.
  In one dimension $\rho_0=1/(\pi v_F)$, therefore 
  $\ell_e\simeq v_F^2/(2{w})$, 
  which coincides with the {\it high energy} (weak disorder) expansion of the 
  localization length 
  $\lambda\simeq2v_F^2/{w}$~\cite{AntPasSly81,LifGrePas88}.
}, 
such a regime does not exist and the system is
either ballistic ($\ell_e\gg L$) or strongly localized ($\ell_e\ll L$).

\vspace{0.25cm}

\noindent
Anderson localization in one dimension has been studied by
mathematicians and mathematical physicists from the view point of spectral
analysis and in connection with limit theorems for products of random
matrices \cite{BouLac85}. A breakthrough was the proof of wave localization 
in one dimension
by Gol'dshtein, Molchanov \& Pastur \cite{GolMolPas77}. Although some
progress has been made, the multidimensional case is still out of reach 
and the subject of weak localization has almost not been touched in the
mathematical literature. One of the main fields of interest in the last twenty
years is the investigation of random Schr\"odinger operators in the presence
of magnetic fields (see for example the 
refs.~\cite{Weg83,BreGroItz84,DesFurOuv95,DesFurOuv96,Fur97,Fur00}~; 
for recent results on Lifshitz tails with magnetic field see 
refs.~\cite{Fur00,LesWar04}).
From the physics side significant advances have been realized. 
Field theoretical methods based on supersymmetry provide a general framework
for disordered systems and also allow establishing some links with quantum
chaos~\cite{Efe97}.

\vspace{0.25cm}

While writing this review, we have tried to collect a large list of references
which is however far from being exhaustive.  Ref.~\cite{LifGrePas88}
provides a reference book on strong localization, mostly focused on spectrum
and localization properties (see also ref.~\cite{Luc92} for a review on 1d
discrete models and Lifshitz tails). A recent text about disorder and random
matrix theory is ref.~\cite{Efe97}.  Many excellent reviews have been written
on weak localization among which \cite{Ber84,ChaSch86} (for the role of
disorder and electron-electron interaction, see the book \cite{EfrPol85} 
or \cite{LeeRam85}). A recent reference is~\cite{AkkMon04}.

\subsection{Overview of the paper}

Section \ref{sec:wl} shows that several functionals of the Brownian motion
arise in the study of electronic transport in weakly disordered metals or
semiconductors (weakly localized). 
The brief presentation of weak localization given in
sections~\ref{sec:brownian1} and~\ref{sec:brownian2} follows the heuristic
discussion of refs.~\cite{AltLee88,ChaSch86}, which is based on the picture
proposed by Khmel'nitzki{\u\i} \& Larkin. In spite of its heuristic
character, it allows drawing suggestive connections with well known
functionals of the Brownian motion.

The sections \ref{sec:expfun} and \ref{sec:evss} deal with problems of strong
localization in one dimension. 
A powerful approach to handle such problems is the phase
formalism\footnote{
  The phase formalism is a continuous version of the Dyson-Schmidt method
  \cite{Dys53,Sch57}. A nice presentation can be found
  in ref.~\cite{Luc92}.
} (presented in refs.~\cite{AntPasSly81,LifGrePas88} for instance). This
formalism leads to a broad variety of stochastic processes.
Section~\ref{sec:expfun} discusses scattering properties of a random potential
and section~\ref{sec:evss} studies spectral properties of a Schr\"odinger
operator defined on a finite interval.

In section \ref{sec:graphs}, we review some results obtained for networks of
wires (graphs), which can be viewed as systems of intermediate dimension
between one and two. We will put the emphasis on spectral determinants, which
appear to be an efficient tool to construct several generating functions
characterizing the diffusion on graphs (or its discrete version, the random
walk).

Finally, in section \ref{sec:magimp}, we show that the physics of a two
dimensional quantum particle submitted to the magnetic field of an assembly of
randomly distributed magnetic vortices involves fine properties of the
planar Brownian motion.


\section{Functionals of the Brownian motion in the context of 
         weak localization\label{sec:wl}}

\subsection{Feynman paths, Brownian motion and weak localization
            \label{sec:brownian1}}

In the path integral formulation of quantum mechanics, each trajectory is
weighted with a phase factor $\EXP{\I S/\hbar}$ where $S$ is the classical
action evaluated along this trajectory. The superposition principle states
that the amplitude of propagation of a particle between two different points
is given by the sum of amplitudes over all paths connecting these two points.
This formulation of quantum mechanics is most useful when there is a small
parameter with respect to which one can make a quasiclassical expansion. One
encounters a similar situation in the derivation of geometrical optics starting
from wave optics. In this case the small parameter is the ratio of the wave
length to the typical distances which are involved in the problem and the
classical paths are light rays. In the context of weak localization the small
parameter is $\lambda_F/\ell_e$ and the quasiclassical approximation amounts
to sum over a certain subset of Brownian paths. The fact that Brownian paths
come into the problem is not so surprising if one goes back to our previous
physical picture of electrons performing random walk due to the scattering by
the impurities. In the continuum limit, describing the physics at length
scales much larger than $\ell_e$, this random walk may be described as a
Brownian motion. It can be shown that $\smean{\Delta\sigma}$ is related to the
time integrated probability for an electron to come back to its initial
position (see ref.~\cite{ChaSch86} for a heuristic derivation or 
ref.~\cite{AkkMon04} for a recent derivation in space representation)~:
%
%
%
\be\label{WL}
\smean{\Delta\sigma}= -\frac{2e^2 D}{\pi\hbar} \int_{\tau_e}^\infty \D t\, 
{\cal P}(\vec r,t|\vec r,0)\,\EXP{-t/\tau_\varphi}
\:.\ee
where $D$ is the diffusion constant and the factor $2$ accounts for spin 
degeneracy. 
${\cal P}(\vec r,t|\vec r\,',0)$ is the Green's function of the diffusion 
equation
\be
\left(\drond{}{t}-D\Delta\right)\,{\cal P}(\vec r,t|\vec r\,',0)
= \delta(\vec r-\vec r\,')\,\delta(t)
\:.\ee
In eq.~(\ref{WL}) the exponential damping at large time describes the lack of
phase coherence due to inelastic processes. The phase coherence time
$\tau_\varphi$ can be related to the phase coherence length by
$L_\varphi^2=D\tau_\varphi$. We have also introduced a cut-off at short time
in eq.~(\ref{WL}), that takes into account the fact that the diffusion
approximation is only valid for times larger than~$\tau_e$.


\subsection{Planar Brownian motion~: stochastic area, winding and
            magnetoconductance\label{sec:brownian2}}

Weak localization corrections are  directly related to the
behaviour of the probability of return to the origin and thus to recurrence
properties of the diffusion process. It therefore follows that dimension $d=2$
plays a very special role. Consider for instance a thin film whose thickness
$a$ is much less than $L_\varphi$. The sample is effectively two dimensional 
and the correction to the conductivity is given by
\be
\smean{\Delta\sigma}= - \frac{e^2}{\pi^2\hbar} \ln(L_\varphi/\ell_e).
\ee
On probabilistic grounds the logarithmic scaling is of course not unexpected
here, it is the same logarithm which occurs in asymptotic laws of the planar
Brownian motion \cite{PitYor86,LeG92}. 
The neighbourhood recurrence of the planar
Brownian motion favors the quantum interference effects and leads to a
reduction of the electrical conductivity. A more striking effect is predicted
if one applies a constant and homogeneous magnetic field ${\cal B}$ over the
sample. In this case the classical action contains a coupling to the magnetic
field $S=e{\cal B}A$ where $A$ is a functional of the path given by the line
integral
\be
A= \frac12 \int (x \D y -y \D x) 
\:.\ee
Properly interpreted, this line integral is nothing but the stochastic area
of planar Brownian motion, whose distribution was
first computed by P.~L\'evy before the discovery of
the Feynman path integral. The weak localization correction reads
\be
\smean{\Delta\sigma({\cal B})}-\smean{\Delta\sigma(0)}= 
\frac{e^2}{2\pi^2\hbar}\int_0^\infty \frac{\D t}{t}\,
\EXP{-t/\tau_\varphi}\,
\left(1-{\rm E}\!\left[\EXP{2\I e{\cal B}A/ \hbar}\right]\right)
\:.\ee
The coupling to the magnetic field now appears with an additional factor $2$
coming from the fact that the weak localization describes quantum
interferences of reversed paths. The expectation ${\rm E}[\cdots]$ is taken
over Brownian loops for a time $t$. The magnetoconductivity is given by
\cite{AltKhmLarLee80,Ber84,ChaSch86}
\be\label{B}
\smean{\Delta\sigma({\cal B})}-\smean{\Delta\sigma(0)}= 
\frac{e^2}{2\pi^2\hbar}\left[
  \psi\left(\frac12+\frac{\phi_0}{8\pi {\cal B} L_\varphi^2} \right)
  - \ln\left(\frac{\phi_0}{8\pi {\cal B} L_\varphi^2} \right)
\right]
\simeq\frac43 \frac{e^2}{\hbar} 
\left(\frac{{\cal B}L_\varphi^2}{\phi_0}\right)^2
\:,\ee
where $\psi(z)$ is the digamma fonction and $\phi_0=h/e$ is the quantum flux.
The {\it r.h.s} corresponds to the weak field limit ${\cal
B}\ll\phi_0/L_\varphi^2$. This increase of the conductivity with magnetic
field is opposite to the behaviour expected classically. This phenomenon is
called ``positive (or anomalous) magnetoconductance''. Experimentally this
effect is of primary importance~: the magnetic field dependence allows one to
distinguish the weak localization correction from other contributions and
permits to extract the phase coherent length $L_\varphi$. This expression fits
the experimental results remarkably well\footnote{
  Note that in materials with strong spin orbit scattering, like gold, the
  effect is reversed and a negative magnetoconductance is observed at small
  magnetic fields~\cite{HikLarNag80,Ber84}. 
} \cite{Ber84}.

Strictly speaking eq.~(\ref{B}) only holds for an infinite sample. In the case
of bounded domains the variance depends on the geometry of the system and can
be computed explictly for rectangles and strips \cite{DesCom92}. The case of
inhomogeneous magnetic fields can be treated along the same line. Consider for
instance a magnetic vortex carrying a flux $\phi$ and threading the sample at
point $0$. In this case the functional which is involved is not the stochastic
area but the index of the Brownian loop with respect to $0$ (winding number).
The probability distribution of the index has been computed independently by
Edwards \cite{Edw67} in the context of polymer physics and Yor \cite{Yor80} in
relation with the Hartman-Watson distribution. For a planar Brownian motion
started at $r$ and conditioned to hit its starting point at time 1, the
distribution of the index when $n$ goes to infinity is
\be
{\rm Proba}\left[{\rm Ind}=n\right] \simeq
\frac{1}{2\pi^2 n^2}\, {\rm K}_0 (r^2)\,\EXP{-r^2}
\:,\ee
where ${\rm K}_0(x)$ is the modified Bessel function of second kind (MacDonald 
function).
The fact that the even moments of this law are infinite is reflected in the
non analytic behaviour of the conductivity
\be\label{eq9}
\smean{\Delta\sigma(\phi)}-\smean{\Delta\sigma(0)}\propto \left|\phi \right|
\:.\ee
It is interesting to
compare with eq.~(\ref{B}) where the quadratic behaviour is given by the
second moment of the stochastic area \cite{Mon95}. The behaviour given
by eq.~(\ref{eq9}) has been observed experimentally \cite{Gei89,BenKliPlo90}
and a theoretical interpretation is provided in ref.~\cite{RamShe87}.

\subsection{Dephasing due to electron-electron interaction and 
                functionals of Brownian bridges}

\noindent
{\it Dephasing in a wire~: relation with the area below a Brownian bridge.--}
Another interesting example of a non trivial connection between a physical
quantity and a functional of the Brownian motion occurs in the study of
electron-electron interaction and weak localization
correction in a quasi-1d metallic wire. 
In ref.~\cite{AltAroKhm82}, Al'tshuler, Aronov \& Khmel'nitzki{\u\i} (AAK)
proposed to model the effect of the interaction between an electron and its
surrounding environment by the interaction with a fluctuating classical field.
The starting point is a path integral representation of the probability
${\cal P}(\vec r,t|\vec r\,',0)$ 
in which is included the effect of the fluctuating 
field that brings a random phase.
After averaging over Gaussian fluctuations of the field, given by the
fluctuation-dissipation theorem, AAK obtained
\be\label{AAK}
\smean{\Delta\sigma}=-2\frac{e^2D}{\pi\sw}
\int_0^\infty\D t\:\EXP{-\gamma t}
  \int_{x(0)=x}^{x(t)=x}{\cal D}x(\tau)\,
  \EXP{-\frac1{4D}\int_0^t\D\tau\,\dot x(\tau)^2
       -\frac{e^2T}{\sigma_0\sw}\int_0^t\D\tau\,|x(\tau)-x(t-\tau)|}
\:,\ee
where $T$ is the temperature (the Planck and Boltzmann constants are 
set equal to unity $\hbar=k_B=1$), and $\sw$ is the area of the section of the 
quasi-1d wire. 
In contrast with eq.~(\ref{WL}) where the loss of phase coherence was
described phenomenologically by an exponential damping\footnote{
  Note that the introduction of an exponential damping describes rigorously 
  several effects~: the loss of phase coherence due to spin-orbit 
  scattering or spin-flip \cite{HikLarNag80}, the penetration of a
  weak perpendicular magnetic field in a quasi 1d wire \cite{AltAro81}.
  In this latter case the effect of the magnetic field is taken into 
  account through $\gamma=\frac13(e{\cal B}W)^2$ where $W$ is the width of the 
  wire of rectangular section.
} with the parameter
$\tau_\varphi$, here the electron-electron interaction affects the weak
localization correction through the introduction in the action of the 
functional of the Brownian bridge\footnote{
  A Brownian bridge, $(x(\tau),\,0\leq\tau\leq1\,|\,x(0)=x(1))$, is a
  Brownian path conditioned to return to its starting point.
}
\be\label{Atilde}
\widetilde{\cal A}_t=\int_0^t\D\tau\,|x(\tau)-x(t-\tau)|
\:.\ee
The additional damping $\exp{-\gamma t}$ in eq.~(\ref{AAK}) describes the 
loss of phase coherence due to other phase breaking mechanisms.
By using a trick which makes the path integral local in time, AAK have 
computed explicitely the path integral and obtained~:
\be\label{AAKres}
\smean{\Delta\sigma}=\frac{e^2}{\pi\sw}\,L_N\,
\frac{{\rm Ai}(\gamma\tau_N)}{{\rm Ai}'(\gamma\tau_N)}
\:,\ee
where ${\rm Ai}(z)$ is the Airy function.
The Nyquist time $\tau_N=(\frac{\sigma_0\sw}{e^2 T\sqrt{D}})^{2/3}$, gives the
time scale over which electron-electron interaction is effective and therefore
plays the role of a phase coherence time. We have also introduced the
corresponding length $L_N=\sqrt{D\tau_N}$. 
Note that the $T$ dependence of $\tau_N\propto T^{-2/3}$ directly reflects the
scaling of the area with time~:  
$\widetilde{\cal A}_t\eqlaw t^{3/2}\widetilde{\cal A}_1$. We stress that the 
AAK theory makes a quantitative prediction for the dependence of the phase
coherence length $L_N$ as a function of the temperature, which has been
verified experimentally for a wide range of parameters~: on metallic (Gold)
wires \cite{EchGerBozBogNil93b,PieGouAntPotEstBir03} 
(the behaviour of $L_N\propto(\sw/T)^{1/3}$ was
observed before in Aluminium and Silver wires \cite{WinRooChaPro86}), and on
wires etched at the interface of two semiconductors \cite{ThoPepAhmAndDav86}.

\begin{figure}[!ht]
\begin{center}
\includegraphics[scale=0.9]{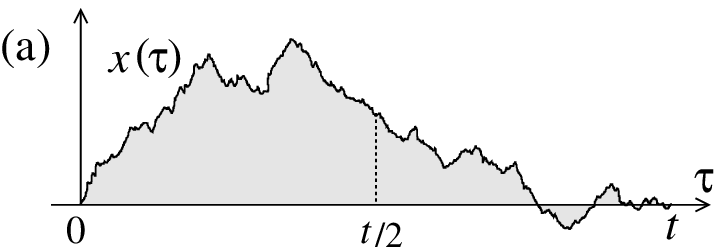}
\hspace{1cm}
\includegraphics[scale=0.9]{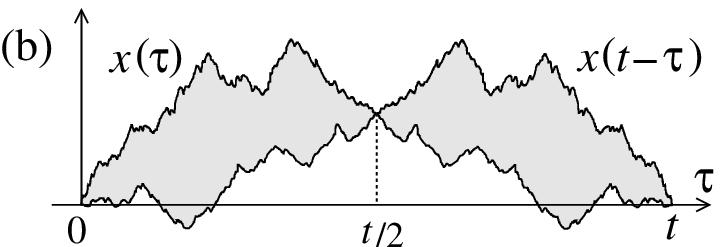}
\end{center}
\caption{
         Left~: {\it The functional ${\cal A}_t$ gives the absolute area 
         below a Brownian bridge $x(\tau)$.}
         Right~: {\it  The functional $\widetilde{\cal A}_t$ measures
         the area between the Brownian bridge $x(\tau)$ and its 
         time reversed counterpart $x(t-\tau)$.
         The two functionals are equal in law [eq.~(\ref{AAtilde})].}
         \label{fig:bridge}}
\end{figure}

It is interesting to point out that the result of AAK can be interpreted as 
the Laplace transform of the distribution of the functional (\ref{Atilde}), 
${\rm E}[\exp{-p \widetilde{\cal A}_t}]$. This functional represents the area
between a Brownian bridge and its time reversed  counterpart ({\it cf.}
figure~\ref{fig:bridge}b), where ${\rm E}[\cdots]$ describes averaging over
Brownian bridges. The conjugate parameter $p$ is played in (\ref{AAK}) by the
temperature~:  $p=\frac{e^2T}{\sigma_0\sw}$.
The result (\ref{AAKres}) has also been derived in the probability 
literature. 
Let us consider the functional
\be\label{A}
{\cal A}_t=\int_0^t\D\tau\,|x(\tau)|
\:,\ee
giving the absolute area below the Brownian bridge starting from the origin,
$x(0)=x(t)=0$ ({\it cf.} figure~\ref{fig:bridge}a). The double Laplace
transform of the distribution of ${\cal A}_t$ has been computed first by
Cifarelli and Regazzini \cite{CifReg75} in the context of economy and
independently by Shepp \cite{She82}, and led to
$
\int_0^\infty\frac{\D t}{\sqrt{t}}\EXP{-\gamma t}
{\rm E}[\EXP{-\sqrt{2}{\cal A}_t}]
=-\sqrt\pi\,\frac{{\rm Ai}(\gamma)}{{\rm Ai}'(\gamma)}
$, 
which is equivalent\footnote{
  Note that the $1/\sqrt{t}$ in the integral computed by Shepp corrects the 
  fact that the averaging over Brownian curves in (\ref{AAK}) is not 
  normalized to unity, whereas ${\rm E}[\cdots]$ is.
} to the result (\ref{AAKres}).
The connection between the two results is clear from the equality in 
law\footnote{
  For a given process $x(\tau)$ the two functionals ${\cal A}_t$ and 
  $\widetilde{\cal A}_t$ are obviously different, however they are 
  distributed according to the same  probability distribution. 
  They are said ``equal in law''.
  \label{footnoteEL}
}~:
\be\label{AAtilde}
{\cal A}_t \eqlaw\widetilde{\cal A}_t
\ee
which follows from\footnote{
  The relation (\ref{anicerel}) is easily proved by using that a Brownian
  bridge $(x(\tau),\,0\leq\tau\leq1\,|\,x(0)=x(1)=0)$ can be written in terms 
  of a free Brownian motion  $(B(\tau),\,\tau\geq0\,|\,B(0)=0)$ as~:
  $x(\tau)=B(\tau)-\tau\,B(1)$.
}
\begin{equation}
 \label{anicerel}
 x(\tau)-x(t-\tau) \eqlaw x(2\tau) 
 \hspace{0.5cm} \mbox{ for } \hspace{0.5cm}
 \tau\in[0,t/2]
 \:.
\end{equation}
The relation (\ref{AAtilde}) allows us to understand more deeply the trick 
used by AAK to make the path integral (\ref{AAK}) local in time.
The distribution of the area ${\cal A}_t$ has been also studied by Rice in 
ref.~\cite{Ric82}. 
Interestingly, the inverse Laplace transform of eq.~(\ref{AAKres}) obtained by
Rice has been rederived recently independently in ref.~\cite{MonAkk05} in 
order to analyze
the loss of phase coherence due to electron-electron interaction in a time
representation.

The study of statistical properties of the absolute area below a Brownian
motion was first addressed by Kac \cite{Kac46} for a free Brownian motion,
long before the case of the Brownian bridge studied by Cifarelli \&
Regazzini and by Shepp. Later on, it has been extended to other
functionals of excursions and meanders\footnote{
  An excursion is a part of a Brownian path between two consecutive zeros
  and a meander is the part of the path after the last zero.
} \cite{JeaPitYor97,PerWel96}, 
which arise in a number of seemingly unrelated problems such as computer 
science, graph theory \cite{FlaLou01}  and statistical physics
\cite{MajCom04,MajCom05}.

\vspace{0.25cm}

\noindent{\it Dephasing in a ring.--}
Very recently, the question of dephasing due to electron-electron interaction
in a weakly disordered metal has been raised again in ref.~\cite{LudMir04}. In 
particular it has been shown that the effect of the geometry
of the ring and the effect of electron-electron interaction combine 
in a nontrivial way leading to a behaviour of the 
harmonics of the magnetoconductance that differs from the one predicted by
eq.~(\ref{WL}) in ref.~\cite{AltAroSpi81}.
For the ring, the functional describing the effect of electron-electron 
interaction on weak localization is now given by
\begin{equation}
  \label{functring}
  {\cal R}_t = \int_0^t\D\tau\,
  |x(\tau)|\, \left( 1 - \frac{|x(\tau)|}{L} \right) 
\end{equation}
instead of the area defined by eq.~(\ref{A}). 
$x(\tau)$ is a Brownian path\footnote{
  As for the case of the wire, the functional involved in the study of
  dephasing in the ring ~\cite{TexMon05b} involves the difference
  $x(\tau)-x(t-\tau)$ instead of the bridge $x(\tau)$. We have used
  the equality in law (\ref{anicerel}) which implies the following property~:
  given a Brownian bridge $(x(\tau),\,0\leq\tau\leq{t}\,|\,x(0)=x(t)=0)$,
  for any even function $f(x)$ we have
  $\int_0^t\D\tau\,f(x(\tau)-x(t-\tau))\eqlaw\int_0^t\D\tau\,f(x(\tau))$.
} on a ring of perimeter $L$ such that $x(0)=x(t)=0$.
The question has been re-examined in more detail in ref.~\cite{TexMon05b},
where the double Laplace transform of the distribution
$
\int_0^\infty\D t\,\EXP{-\gamma t}\,
\frac{1}{\sqrt{t}}\EXP{-\frac{(nL)^2}{2t}}\,{\rm E}_n[\EXP{-{\cal R}_t}]
$
has been derived. ${\rm E}_n[\cdots]$ denotes averaging over 
Brownian bridges defined on a circle, with winding number $n$. Note that the 
Laplace transform of the distribution ${\rm E}_n[\EXP{-{\cal R}_t}]$ 
has been also studied in ref.~\cite{TexMon05b}.


\section{Exponential functionals of Brownian motion and Wigner time delay
         \label{sec:expfun}}

\subsection{Historical perspective}

Exponential functionals of the form,
\begin{equation}
\label{deffe}
A_t^{(\mu)}= \int_0^t \D s\, \exp{-2(B(s) +\mu s)}
\end{equation}
where $(B(s),\,s\geq0,\,B(0)=0)$ is an ordinary Brownian motion have been the 
object of many studies in mathematics \cite{Yor00}, mathematical finance and 
physics \cite{ComMonYor98}. 
In the physics of classical disordered systems, the starting point was the
analysis of the series (Kesten variable)
\be
Z= z_1 +z_1z_2+z_1z_2z_3+\cdots
\ee
where the $z_i$ are independent and identical random variables.
Several papers have been devoted to the study of this random variable in the
physics \cite{CalLucNiePet85,DerHil83} and mathematics \cite{Kes73,Ver79}
literature.
It was  realized that $Z$ is the discretized version of $ A_\infty^{(\mu)}$
defined in eq.~(\ref{deffe}),
which may be interpreted as a trapping time in the context of classical
diffusion in a random medium. The tail of the
probability distribution $P(Z)$ controls the anomalous diffusive behaviour of a
particle moving in a one-dimensional random force field \cite{BouComGeoLeD90}. 
The functional
$A_t^{(\mu)}$ also arises in the study of the transport properties of
disordered samples of finite length \cite{MonCom94,OshMogMor93}.
In the context of one dimensional localization the fact that the norm of the
wave function is distributed\footnote{
  When $\mu>0$, $A_t^{(\mu)}$ possesses a limit distribution for $t\to\infty$.
  Moreover, we have the equality in law (see footnote \ref{footnoteEL})~:
  $A_\infty^{(\mu)}\eqlaw 1/\gamma^{(\mu)}$, where 
  $\gamma^{(\mu)}$ is distributed according to a $\Gamma$-law~:
  $P(\gamma)=\frac1{\Gamma(\mu)}\gamma^{\mu-1}\EXP{-\gamma}$.
  \label{foot:gammalaw}
} as $ A_\infty^{(1)}$ is mentioned in the book of Lifshitz {\it et al}
\cite{LifGrePas88}.

\subsection{Wigner time delay\label{sec:wtd}}

As discussed in the introduction, the localization of quantum states in one
dimension is well understood. However since real systems are not infinite,
asking about the nature of the states which are not in the bulk is a perfectly
legitimate question. It was pointed out by Azbel in ref.~\cite{Azb83} 
that, when a disordered region is connected to a region free of disorder,
the localized states acquire a finite lifetime which
shows up in transport through sharp resonances,
refered as ``Azbel resonances''. This picture was used in
ref.~\cite{JayVijKum89} where it was argued that these resonances could be
probed in scattering experiments and lead to an energy dependent random time
delay of the incident electronic wave. Motivated by this result and by
developments in random matrix theory we consider the scattering problem for
the one dimensional Schr\"odinger equation (in units $\hbar=2m=1$)
\be
-\frac{\D^2}{\D x^2}\psi(x)+V(x)\psi(x)=E\,\psi(x)
\ee
defined on the half line $x\geq 0$  with the Dirichlet boundary conditions
$\psi(0)=0$.
The potential $V(x)$ has its support on the interval $[0,L]$.
Outside of this interval, the scattering state of energy $E=k^2$ is given by
\be\label{scattstate}
  \psi_E(x)=\frac1{\sqrt{hv_E}}
  \left( \EXP{-\I k(x-L)} +\EXP{\I k(x-L)+\I\delta} \right)
\:,\ee 
where $h=2\pi$ is the Planck constant (in unit $\hbar=1$) and 
$v_E=\D E/\D k=2k$ the group velocity.
Eq.~(\ref{scattstate})
represents the superposition of an incoming plane wave incident from
the right and a reflected
plane wave caracterized by its phase shift $\delta(E)$. 
The Wigner time delay, defined by the relation $\tau=\D\delta/\D{E}$,
can be understood as the time spent by the wave packet of energy $E$ 
in the disordered region (this interpretation is only valid 
at high energy~; see 
refs.~\cite{HauSto89,But90,LanMar94,CarNus02} for review articles on time 
delay and traversal times).
This representation of the time delay has been used in many papers both
in the mathematics and physics literature. Starting from this representation
one can derive a system of stochastic differential equations which can be
studied in certain limiting cases. However, to understand the universality 
of the statistical properties of the Wigner time in the high energy limit,
we follow a different approach below.

\subsubsection*{Universality of time delay distribution at high energy}

Using the expression of the scattering state (\ref{scattstate}) and the
definition $\tau=\D\delta/\D{E}$, we can obtain the so-called Smith formula
\cite{Fri58,Smi60} relating the time delay to the wave function in the bulk~:
\be\label{Smith}
  \tau(E)= 2\pi\int_0^L \D x\, |\psi_E(x)|^2 
  -\frac{1}{2E}\sin\delta(E)
\:.\ee
Following
ref.~\cite{AntPasSly81} one can parametrize the wave function in the bulk in
terms of its phase and modulus. 
For this purpose we rewrite the Schr\"odinger equation as a set of two 
coupled first order differential equations, $\D\psi_E/\D x=\psi'_E$
and $\D\psi'_E/\D x=(V(x)-E)\psi_E$, and perform the change of variables~:
$\psi_E(x)=\EXP{\xi(x)}\sin\theta(x)$ and 
$\psi'_E(x)=k\EXP{\xi(x)}\cos\theta(x)$. One obtains a new set of 
first order differential equations $\D\theta/\D x=k-\frac1kV(x)\sin^2\theta$
and $\D\xi/\D x=\frac1{2k}V(x)\sin2\theta$. Up till now everything is exact. 
If we now consider the high energy limit\footnote{
  To define precisely the high energy limit, let us introduce the integral
  of the correlation function of the disorder
  ${w}=\int\D x\,\smean{V(x)V(0)}$. The high energy limit corresponds to
  $k\gg{w}^{1/3}$.
}
we can neglect the second term on the {\it r.h.s} of eq.~(\ref{Smith})
and integrate out the phase which is a fast variable~; we obtain
\be\label{timedelay0}
\tau = \frac{1}{k}\int_0^L \D x \:\EXP{2(\xi(x)-\xi(0))}
\:.
\ee
This representation of the time delay holds for any realization of the
disordered potential. Moreover, one can prove under rather mild conditions
on the correlations of the random potential
that $\xi(x)$ is a Brownian motion with drift 
$\xi(x)=x/\lambda+\sqrt{1/\lambda}\,B(x)$ where
$B(x)$ is an ordinary Brownian motion and $\lambda$ the localization length. 
Using the scaling properties of the Brownian motion gives the following
identity in law
\be\label{timedelay}
\tau \eqlaw \frac{\lambda}{k}\int_0^{L/\lambda} \D x\: \EXP{-2(B(x)+x)}
= \frac{\lambda}{k}\, A_{L/\lambda}^{(1)}
\:,
\ee
where $A_L^{(1)}$ has been defined in eq.~(\ref{deffe}).
This representation of the time delay as an exponential functional of the
Brownian motion, first established in ref.~\cite{FarTsa94} by a different
method, allows us to obtain a number of interesting results
\cite{ComTex97,TexCom99}~:

\vspace{0.25cm}

\noindent
({\it i}) Existence of a limit distribution for fixed\footnote{
  From the remark of the footnote \ref{foot:gammalaw}, we notice that 
  $1/\tau$ is distributed according to an exponential law.
} $\tau$ and $ L\rightarrow\infty$
\be
P(\tau)=\frac{\lambda}{2k\tau^2}\, \EXP{- \frac{\lambda}{2k\tau}} 
\:.\ee
This result is reminiscent of the random matrix theory prediction in
spite of the fact that this theory does not apply to systems that are
strictly one-dimensional\footnote{
  The random matrix theory describes the regime of weak localization or 
  systems whose classical dynamics is chaotic. The distribution of 
  time delay in this framework has the same functional form 
  $P(\tau)\propto\frac1{\tau^{2+\mu}}\EXP{-\tau_0/\tau}$ 
  \cite{FyoSom96a,GopMelBut96,BroFraBee97,OssFyo05} with two differences~: 
  ({\it i}) the exponent $2+\mu>2$,
  ({\it ii}) the time scale $\tau_0$.
}. 

\vspace{0.25cm}

\noindent
({\it ii}) Linear divergence of the first moment, $\smean{\tau}=L/k$, and 
exponential divergence of the higher moments 
$\smean{\tau^n}\propto\EXP{2n(n-1)L/\lambda}$.
This divergence reflects a log-normal tail of the distribution for a finite
length $L$.

\subsubsection*{Time delay and density of states}

The relation (\ref{Smith}) also provides another interpretation to these 
results. In the situation under consideration, the scattering state is 
directly related to the local density of states (LDoS) by 
$\rho(x;E)=\bra{x}\delta(E-H)\ket{x}=|\psi_E(x)|^2$. Therefore the time delay
can be interpreted as the DoS of the disordered region\footnote{
  The relation between time delay and DoS is sometimes refered as Krein-Friedel
  relation \cite{Fri52,Kre53}. This relation has been originally introduced in
  ref.~\cite{BetUhl37} in the context of statistical physics 
  (see also ref.~\cite{DasMaBer69} and \S77 of ref.~\cite{LanLif66e}). 
  Note that it has been recently
  discussed in the context of graphs in refs.~\cite{Tex02,TexBut03,TexDeg03}.
}. 
This establishes a relation between the results given above and the work of
Al'tshuler \& Prigodin~\cite{AltPri89} where the distribution of the LDoS
was studied for a white noise potential using the method of
Berezinski{\u\i}\footnote{
  The method of Berezinski{\u\i} blocks has been introduced to study
  specifically the case of white noise disordered potential in 
  one dimension \cite{Ber74}.
  This powerful approach has been widely used and has allowed to derive
  numbers of important results like in refs.~\cite{GogMelRas76,GorDorPri83} 
  for example.
}.

\subsubsection*{Time delay for Dirac Hamiltonian at the threshold energy}

The representation (\ref{timedelay}) of the time delay as an exponential
functional of the Brownian motion only holds in the weak disorder ({\it i.e.}
high energy) limit. A similar representation can be derived for the random
mass Dirac model at the middle of the spectrum. 
The Dirac Hamiltonian~is 
\be\label{HDirac}
H_{\rm D}=\sigma_2\,\I\frac{\D}{\D x} + \sigma_1\phi(x)
\:,\ee
where $\sigma_i$ are the Pauli matrices. $\phi(x)$ can be interpreted as a
mass\footnote{
  The Hamiltonian (\ref{HDirac}) with $\phi(x)$ a white noise was introduced
  by Ovchinnikov \& Erikmann \cite{OvcEri77} as a model of 
  one-dimensional semiconductor with a narrow fluctuating gap.
  It is interesting to point out that the Dirac equation $H_{\rm D}\psi=k\psi$ 
  also has an interpretation in the context of superconductivity, as 
  linearized Bogoliubov-de~Gennes equations for a real random superconducting
  gap $\phi(x)$.
  Finally it is worth mentioning that more general 1d Dirac Hamiltonians 
  with several kinds of disorder (mass term, potential term and magnetic 
  field) have been studied in refs.~\cite{Boc99,Boc00}. 
}. 
The Dirac equation $H_{\rm D}\psi=k\,\psi$ possesses a
particle-hole symmetry reflected in the symmetry of the spectrum with respect
to $k=0$ (the dispersion relation of the Dirac equation is linear 
in absence of the mass term and energy is equal to momentum).
The middle of the spectrum is an interesting point where the divergence in the
localization length signals the existence of a delocalized state
(see footnote \ref{footnoteSUSY}). These
properties should therefore show up in the probability distribution of the 
Wigner time delay. 
Since the dispersion relation of the Dirac equation is linear, the time 
delay is now defined as $\tau=\D\delta/\D k$.
The following representation has been derived 
in ref.~\cite{SteCheFabGog99}~:
\be \label{tdsusy}
\tau= 2\int_0^L \D x\, \EXP{2\int_0^x \phi(y)\D y}
\:.\ee
In contrast with
eqs.~(\ref{timedelay0},\ref{timedelay}), which have been obtained after
averaging over the fast phase variable, the representation (\ref{tdsusy}) is
exact (for $k=0$). If $\phi(x)$ is a white noise, then
\begin{equation}
\label{ChenTD}
\tau = \frac{2}{g}\,A_{g L}^{(0)}
\:.\end{equation}
In this case since the drift vanishes, it is known \cite{MonCom94,Yor00} that
there is no limiting distribution when $L\rightarrow\infty$. The physical
meaning of the lognormal tail
\be
P(\tau)\sim \frac{1}{2\tau\sqrt{2\pi gL}}\, \EXP{-\frac{1}{8gL} \ln^2(g\tau)}
\ee
in the limit $\tau\rightarrow\infty$ is discussed in detail in
ref.~\cite{SteCheFabGog99}.


\section{Extreme value spectral statistics\label{sec:evss}}

The density of states (DoS) and the localization length of a one-dimensional
Hamiltonian with a random potential can be studied by the phase formalism
\cite{AntPasSly81,LifGrePas88} (see also ref.~\cite{Luc92} for discrete
models).
In the previous section we pointed out
that the physics of the strong localization in one-dimension can also be
probed from a different angle by considering the scattering of a plane wave by
the random potential. This has led to the study of the Wigner time delay. In
this section we show that the phase formalism also allows describing finer
properties of the spectrum such as the probability distribution of the $n$-th
eigenvalue. This is an instance of an {\it extreme value statistics} ~: given
a ranked sequence of ${\cal N}$ random variables
$x_1 \leq x_2 \leq \ldots\leq x_{\cal N}$, 
the problem is to find the distribution of the $n$-th of these variables in a
given interval. In the particular case of uncorrelated random variables, the
extreme value distributions have been studied by E.~Gumbel
\cite{Gum35,Gum54,Gum58}. This problem becomes much more complicated when the
random variables are correlated, a case which has recently attracted a lot of
attention. Such extreme value problems have appeared recently in a variety of
problems ranging from disordered systems \cite{BouMez97,CarLeD01,DeaMaj01} to
certain computer science problems such as growing search trees \cite{MajKra03}.
In the case of the spectrum of a random Hamiltonian, the eigenvalues are in
general correlated variables, apart from the case of strongly localized
eigenstates \cite{Mol81}. Below we show that, in the one dimensional case,
this problem is related to studying the first exit time distribution of a 
one-dimensional diffusion process.

\mathversion{bold}
\subsection{Distribution of the $n$-th eigenvalue~: relation with a first 
            exit time distribution}
\mathversion{normal}

We consider a Schr\"odinger equation $H\varphi(x)=E\varphi(x)$ on a finite
interval $[0,L]$. The spectral (Sturm-Liouville) problem is further defined by
imposing suitable boundary conditions. We choose the Dirichlet boundary
conditions $\varphi(0)=\varphi(L)=0$. The spectrum of $H$ is denoted by 
${\rm Spec}(H)=\{E_0<E_1<E_2<\cdots\}$. Our purpose is to compute the
probability 
\be
W_n(E) = \smean{\delta(E-E_n)}
\ee
for the eigenvalue $E_n$ to be at energy $E$ (the bracket $\smean{\cdots}$
means averaging over the random potential).
Note that the sum of these distributions 
$\frac1L\sum_{n}W_n(E)=\rho(E)$ is the average DoS per unit length.
We now show how the calculation of $W_n(E)$ can be cast into  a first exit time
problem.

\subsubsection*{Random Schr\"odinger Hamiltonian}

We consider the Hamiltonian 
\be\label{hamil}
H=-\frac{\D^2}{\D x^2} + V(x)
\:,\ee
where $V(x)$ is a Gaussian white noise random potential~: $\smean{V(x)}=0$ and
$\smean{V(x)V(x')}={w}\,\delta(x-x')$. 

We replace the Sturm-Liouville problem by a Cauchy problem~:
let  $\psi(x;E)$ be the solution of the Schr\"odinger equation
$H\psi(x;E) = E\psi(x;E)$ with the boundary conditions $\psi(0;E)=0$
and $\frac{\D}{\D x}\psi(0;E)=1$. 
The boundary condition $\psi(L;E)=0$ is fulfilled whenever the energy $E$ 
coincides with an eigenvalue $E_n$ of the Hamiltonian. In this case,
the wave function 
$\varphi_n(x)=\psi(x;E_n)/\left[\int_0^L\D{x'}\,\psi(x';E_n)^2\right]^{1/2}$ 
has $n$ nodes in the interval $]0,L[$, and two nodes at the boundaries. 

\begin{figure}[!ht]
\hspace{3cm}
\begin{tabular}{ll}
\underline{$E<E_0$}           &  \diagram{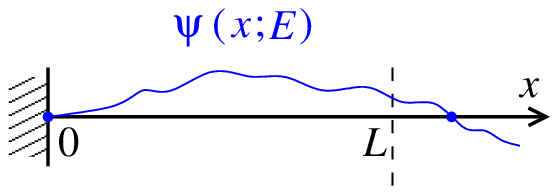}{1}{-0.75cm}  
\\
\underline{$E=E_0$}           &  \diagram{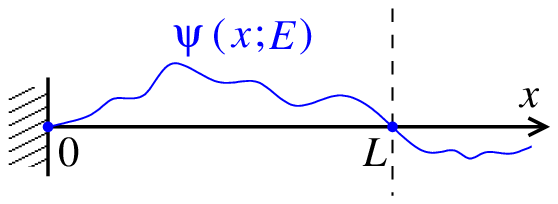}{1}{-0.75cm}  
\\
\underline{$E_0<E<E_1$}       &  \diagram{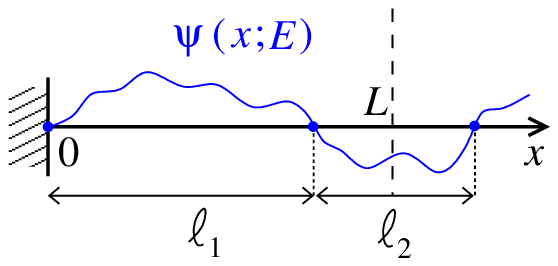}{1}{-1.40cm}  
\\[0.5cm]
\hspace{1.5cm}{\huge$\vdots$} & \hspace{1cm}{\huge$\vdots$\hspace{2cm}$\vdots$}
\end{tabular}
\caption{\it
         The probability for the $n$-th level $E_n$ of the spectrum to be 
         at $E$ is also the probability for the $n+1$-th node of $\psi(x;E)$ 
         to be at $L$.\label{fig:psideE}}
\end{figure}

Let us denote by $\ell_m$ ($m\geq1$), the length between two consecutive nodes.
We consider the Ricatti variable
\be
z(x;E) = \frac{\D}{\D x} \ln|\psi(x;E)|
\:,\ee
which obeys the following equation:
\be\label{Riceq}
\frac{\D}{\D x} z = -E - z^2 + V(x)
\:,\ee
with initial condition $z(0;E)=+\infty$.
This equation may be viewed as a Langevin equation for a particle located at
$z$ submitted to a force $-\partial U(z)/\partial z$ deriving from the 
unbounded potential 
\be\label{potU}
U(z)=Ez+\frac{z^3}{3}
\ee 
and to a random white noise $V(x)$. 

Each node of the wave function corresponds to $|z(x)|=\infty$. At ``time''
$x=0$ the ``particle'' starts from $z(0)=+\infty$ and eventually ends at
$z(\ell_1-0^+)=-\infty$ after a ``time'' $\ell_1$. Just after the first node
it then starts again from  $z(\ell_1+0^+)=+\infty$, due to the continuity of
the wave function. It follows from this picture that the distance
$\ell_m$ between two consecutive nodes may be viewed as the ``time'' needed
by the particle to go through the interval $]-\infty,+\infty[$ (the
``particle'' is emitted from $z=+\infty$ at initial ``time'' and absorbed
when it reaches $z=-\infty$).
The distances $\ell_m$ are random variables, interpreted as times needed
by the process $z$ to go from $+\infty$ to $-\infty$.
These random variables are statistically independent because each time the
variable $z$ reaches $-\infty$, it loses the memory of
its earlier history since it is brought back to the same initial condition
and $V(x)$ is $\delta$-correlated. This remark is a crucial point for the
derivation of $W_n(E)$.
Interest in these random ``times'' $\ell_m$ lies in their relation
with the distribution of the eigenvalues. Indeed the probability that
the energy $E_n$ of the $n$-th excited state is at $E$ is also
the probability that the sum of the $n+1$ distances between the nodes is
equal to the length of the system: $L=\sum_{m=1}^{n+1}\ell_m$
(this is illustrated on figure~\ref{fig:psideE}).
Since the $\ell_m$ are independent and identically distributed
random variables   
$\proba\left[L=\sum_{m=1}^{n+1}\ell_m\right]$ is readily obtained from the
distribution $P(\ell)$ of one of these variables. 

We introduce the intermediate variable ${\cal L}(z)$ giving the time needed by
the process starting at $z$ to reach $-\infty$~: therefore we have 
$\ell={\cal L}(+\infty)$. The Laplace transform of the distribution
$
h(\alpha,z) = 
\smean{ \EXP{-\alpha{\cal L}} \:|\: z(0)=z;\:z({\cal L})=-\infty }
$ 
obeys \cite{Gar89}~:
\be
G_z \, h(\alpha,z) = \alpha \, h(\alpha,z)
\:,\ee
where the backward Fokker-Planck generator is
\be\label{bfpegen}
G_z = -U'(z) \partial_z + \frac{{w}}{2} \partial_z^2
\:.\ee
The boundary conditions are
$\partial_z h(\alpha,z)|_{z=+\infty}=0$ and $h(\alpha,-\infty)=1$.
By Laplace inversion we can derive $P(\ell)$ since
$h(\alpha,+\infty)=\int_0^\infty\D\ell\,P(\ell)\EXP{-\alpha\ell}$.

\begin{figure}[!ht]
\begin{center}
\includegraphics{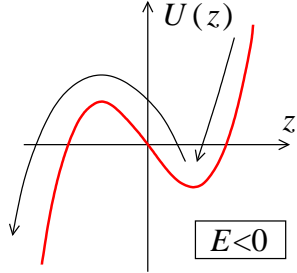}
\end{center}
\caption{\it The potential (\ref{potU}) related to the deterministic force 
         to which the Ricatti variable is submitted in eq.~(\ref{Riceq}).
         \label{fig:potUneg}}
\end{figure}

The solution of this problem is given in ref.~\cite{Tex00}.
Here we only consider the limit $E\to-\infty$, which corresponds to 
the bottom of the spectrum. 
In this regime the dynamics of the Ricatti variable $z$ can be read off 
from the shape of the potential (see figure~\ref{fig:potUneg}).

The particle falls rapidly in the well from which it can only escape by a
fluctuation of the random force. Therefore the time needed to go from
$+\infty$ to $-\infty$ is dominated by the time spent in the well, which is
distributed according to the Arrhenius formula. It follows that the average
``time'', which is related to the inverse of the integrated density of states
(IDoS) per unit length $N(E)=\int_0^E\D E'\,\rho(E')$, behaves as
$
\mean{\ell} = N(E)^{-1}
\simeq \frac{\pi}{\sqrt{-E}} \exp\frac{8}{3{w}}(-E)^{3/2}
$.
This simple picture, first provided by Jona-Lasinio \cite{Jon83} allows
recovering the exponential tail obtained by several methods in 
refs.~\cite{FriLlo60,Hal65,AntPasSly81,LifGrePas88,ItzDro89}\footnote{
  It is interesting to quote ref.~\cite{KucSad98} where it was shown that 
  this non perturbative exponential tail can be obtained by a simple counting
  method of ``skeleton'' diagrams assuming that they all have the same value
  (this approach was also applied to the 3d case).
}.
The distribution of the length $\ell$ is a Poisson law (see \cite{Gar89}
or appendix of ref.~\cite{Tex00}):
\be\label{dol-}
P(\ell) = N(E)\,\exp{-\ell\,N(E)}
\:.\ee
Using this result we can show that \cite{Tex00} 
\be\label{TheResult}
W_n(E) = L\rho(E) \, \frac{(L\,N(E))^n}{n!} \, \EXP{-L\,N(E)} 
\:.\ee
This result has a clear meaning~: $L\rho(E)$ gives the probability 
to find any level at $E$ and the factor $\frac{x^n}{n!}\EXP{-x}$ ``compels''
the number of states below $E$, $x=LN(E)$, to be close to $n$.
We may go further and write
\be
W_n(E) = \frac{1}{\delta E_n} \ 
         \omega_n\left(\frac{E - E_n^{\typ}}{\delta E_n}\right)
\:,\ee
where the typical value of the energy is 
$E_n^{\typ}(L)=-(\frac{3{w}}{8}\ln\tilde L)^{2/3}$,
while the scale of the fluctuations reads 
$
\delta E_n =\frac{ {w}^{2/3} }{2\sqrt{n+1}}(3\ln\tilde L)^{-1/3}
$, 
where $\tilde L=\frac{L\,{w}^{1/3}}{2\pi(n+1)}$.
The function
\be\label{TheResultbis}
  \omega_n(X)=\frac{(n+1)^{n+\frac12}}{n!}
  \exp{\left(\sqrt{n+1}\:X -(n+1)\,\EXP{{X}/{\sqrt{n+1}}}\right)}
\:\ee
has the form of a Gumbel law for {\it uncorrelated} random variables.
The fact that the eigenvalues are uncorrelated is a consequence of the 
strong localization of the eigenfunctions \cite{Mol81}.

The work summarized in the above paragraph, which appeared in
ref.~\cite{Tex00}, generalizes the result of McKean \cite{McK94} for the
ground state.

A similar result was obtained by Grenkova {\it et al} \cite{GreMolSud83} for
the model of $\delta$-impurities with random positions, in the limit of low
impurity density which has no counterpart in the model we considered 
here\footnote{
  See footnote \ref{FrischLloyd}.
}.
However both models describe the same physics of strongly localized
eigenstates.

\subsubsection*{Supersymmetric random Hamiltonian}

Random Schr\"odinger operators have been investigated through a wide range of
models. Depending on the physical context, there are indeed many ways to model 
the
disorder. In the previous paragraph we have assumed that the potential $V(x)$
is a white noise. We have also mentioned that if the potential is a 
superposition of $\delta$-potentials randomly distributed along the line,
although the spectra of the two models are quite different\footnote{
  The model of $\delta$-impurities with random positions was introduced 
  and studied by Schmidt \cite{Sch57} but often refered as the 
  Frisch \& Lloyd model \cite{FriLlo60}. Compared to the model for 
  white noise potential, which is characterized by one parameter, the Frisch \&
  Lloyd model is characterized by two parameters~: the 
  strength of the $\delta$-potential, and their density.
  In the limit of high density of impurities this model is equivalent to 
  the white noise potential model. The limit of low density presents 
  different spectral singularities (Lifshitz singularity and, for 
  negative weight of $\delta$-potentials, an additional Halperin singularity
  in the negative part of the spectrum).\label{FrischLloyd}
}, 
the extreme spectral statistics are the same. The localization properties and
the statistics of the time delay at high energy are also similar for both
models \cite{Tex99,TexCom99}. These two models belong to the same class of
random Hamiltonians with a random scalar potential with short range
correlations. They are both continuous versions of discrete
tight binding models with on-site random potential. This is the case of
so-called {\it diagonal disorder}, since the random potential appears on the
diagonal matrix elements of the tight binding Hamiltonian in the basis of
localized orbitals\footnote{
  Note that the continuum limit of tight binding Hamiltonian with diagonal 
  disorder can lead to different continous models. 
  Let us consider the discrete model
  $H_{i,j} = -\delta_{i,j+1}-\delta_{i,j-1}+\delta_{i,j}V_i$. 
  For $V_i=0$ the spectrum is $E_{k}=-2\cos(k)$, with $k\in]-\pi,\pi]$.
  (A) If the continuous limit is taken by considering the band 
  edge ($k\sim0$), one is led to the continuous model (\ref{hamil}).
  (B) If the band center is considered instead ($k\sim\pi/2$), the 
  spectrum can be linearized and one is led to a Dirac Hamiltonian,
  like in refs.~\cite{Kel64,OvcEri77}.
  This point has been recently rediscussed in ref.~\cite{SchTit03}.
  \label{footnoteDD}
}.

Other interesting models can be constructed by introducing disorder in the
hoppings, instead. We refer to such models as {\it off-diagonal disorder}. The
Dirac Hamiltonian (\ref{HDirac}) introduced above provides a continuum limit
of such a model\footnote{
  The continuum limit of a tight binding Hamiltonian with random hoppings
  has been discussed in ref.~\cite{TakLinMak80} (see also the review in 
  ref.~\cite{Mon95}).
  As pointed in footnote \ref{footnoteDD}, random Dirac Hamiltonians 
  can also appear as continuum limit of the band center of a 
  discrete models with diagonal disorder. 
}.
Dirac Hamiltonians appear naturally in several contexts of condensed matter
physics. The existence of symmetries in the Dirac Hamiltonian\footnote{
  A complete classification of symmetries of disordered Hamiltonians 
  extending the famous Wigner-Dyson ensembles of random matrix theory has 
  been provided in refs.~\cite{Ver94,Zir96,AltZir97}
  (see also the recent review article \cite{HeiHucZir05}). 
}  
(particle-hole, chiral,...) can lead to interesting features in the presence
of disorder, which has attracted some attention
(see for example
refs.~\cite{NerTsvWen94,AltZir97,BroMudFur00,BroMudFur03,BocSerZir00,Boc00,EvaKat03},
the last of these references gives a brief overview).
The square of the Dirac Hamiltonian, 
$H_{\rm D}^2=-{\D_x^2} + \phi(x)^2 +\sigma_3 \phi'(x)$,
is related to the pair of supersymmetric Schr\"odinger isospectral 
Hamiltonians $H_\pm=-{\D_x^2} + \phi(x)^2 \pm \phi'(x)$. When 
$\mean{\phi(x)}=0$ we may forget the sign and simply consider~: 
\be\label{Hsusy}
H_S=-\frac{\D^2}{\D x^2} + \phi(x)^2 + \phi'(x)
\:.\ee
Such Hamiltonians appear in a variety of problems ranging from
the 1d classical diffusion in a random force \cite{BouComGeoLeD90,IglMon05},
electronic structure of polyacetylene \cite{TakLinMak80}, 
spin Peierls chains \cite{FabMel97,FabMel97b,SteFabGog98} and also in a 
continuum limit of the random field XY model \cite{GurCha03}
(see ref.~\cite{ComTex98} for a short review on supersymmetric disordered
quantum mechanics). Spectral and localization properties 
have been studied in detail when $\phi(x)$ is a white noise 
\cite{OvcEri77,LifGrePas88,BouComGeoLeD87,BouComGeoLeD90} and also
when $\phi(x)$ has a finite correlation length as in a random telegraph process
\cite{ComDesMon95} (this last case has found some application in the context
of spin Peierls chains).

In the high energy limit, the localization properties and the statistics of
the time delay do not show any difference with respect to the case of diagonal
disorder. However the low energy properties are quite different. This is
easily understood by noticing that, due to its relation to the Dirac 
Hamiltonian, the supersymmetric Hamiltonian can be
factorized as $H_S=Q^\dagger Q$ where $Q=-\D_x+\phi(x)$ and
$Q^\dagger=\D_x+\phi(x)$. In particular, such a structure enforces a positive
spectrum~:  ${\rm Spec}(H_S)\subset\RR^+$. 
When $\phi(x)$ is white noise of zero mean, $\smean{\phi(x)}=0$ and
$\smean{\phi(x)\phi(x')}=g\,\delta(x-x')$, the case considered below, the
DoS presents a logarithmic Dyson singularity~: the integrated density 
of states (IDoS) reads
$N(E)\sim1/\ln^2E$ \cite{OvcEri77,BouComGeoLeD87,BouComGeoLeD90}, 
while the localization length diverges logarithmically\footnote{
  A more complete picture of the properties of this model at $E=0$ is given in
  several works~:
  (A) moments and correlations of the zero mode are studied in
  refs.~\cite{SheTsv98,ComTex98}. 
  (B) Additionally to the statistical properties of the time delay 
  , eq.~(\ref{ChenTD}),
  (C) the distribution of the transmission probability is derived in
  ref.~\cite{SteCheFabGog99}. In particular it was shown that the 
  average transmission through an interval of length $L$ decreases like 
  $\smean{T}\propto1/\sqrt{L}$, which is slower than in the diffusive regime.
  (D) The existence of a finite conductivity was
  demonstrated in ref.~\cite{GogMel77}.
  \label{footnoteSUSY}
} $\lambda(E)\sim\ln(1/E)$.

It is interesting to investigate the extreme value statistics of the spectrum
in the low energy regime, where we expect properties quite different from the
one obtained for the diagonal disorder. The derivation follows closely 
that in the previous case, however the relevant 
random processes and the approximations are 
different. The first step is to decouple the Schr\"odinger equation
$H_S\varphi(x)=k^2\varphi(x)$ into two first order differential equations 
({\it i.e.} go back to the Dirac equation)~:
\bea
Q^\dagger \chi(x) &=& k \varphi(x) \\
Q \varphi(x)      &=& k \chi(x)
\:.\eea
Then we may use the phase formalism by introducing a phase variable
and an envelope variable~: 
$\varphi(x)=\EXP{\xi(x)} \sin\vartheta(x)$ and 
$\chi(x)=-\EXP{\xi(x)} \cos\vartheta(x)$.
The phase variable obeys a stochastic differential equation with 
a noise multiplying a trigonometric function of the phase.
For convenience we introduce an additive process $\zeta(x)$ defined
as $\zeta(x)=\pm\frac12\ln|\tan\vartheta(x)|$.
The sign $+$ is chosen for 
$(\vartheta\ {\rm mod}\ \pi)\in[0,\pi/2]$ and the sign $-$ for 
$(\vartheta\ {\rm mod}\ \pi)\in[\pi/2,\pi]$. This new process obeys 
the stochastic differential equation
\be \label{eqzeta}
\frac{\D}{\D x}\zeta = k\cosh2\zeta \pm \phi(x)
\:.\ee
Between two nodes of the wave function, the variable $\zeta$ crosses twice the
interval $]-\infty,+\infty[$. Note that when $\phi(x)$ is a white noise of
zero mean, the sign $\pm$ can be disregarded. The study of the time required
by the process to cross the interval can be performed by the same method as
above~:  we introduce the ``time'' $\tilde\Lambda(\zeta)$ needed 
to go from $\zeta$ to $+\infty$.
The Laplace transform of the distribution
$h(\alpha,\zeta) = 
\smean{ \EXP{-\alpha\tilde\Lambda} \:|\:\zeta(0)=\zeta;\ \zeta(\tilde\Lambda)
=+\infty }$
obeys a diffusion equation 
$
(k\cosh2\zeta\:\partial_\zeta + \frac{g}{2}\partial_\zeta^2)
h(\alpha,\zeta) = \alpha \: h(\alpha,\zeta)
$
that involves the backward Fokker-Planck operator related to the 
Langevin equation (\ref{eqzeta}).

\begin{figure}[!ht]
\begin{center}
\includegraphics{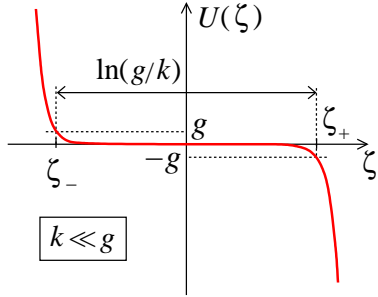}
\end{center}
\caption{\it The potential related to the deterministic force in 
         eq.~(\ref{eqzeta}) felt by the  
         variable $\zeta$.\label{fig:potsusy}}
\end{figure}

In the low energy limit $k\ll g$ we expect that most of the ``time''
$\Lambda\equiv\tilde\Lambda(-\infty)$ is spent in the region where the
potential is almost flat (see figure~\ref{fig:potsusy}), therefore we replace
the diffusion equation for $h(\alpha,\zeta)$ by the free diffusion equation
$
\frac{g}{2}\partial_\zeta^2h(\alpha,\zeta) = \alpha \: h(\alpha,\zeta)
$ 
on the finite interval $\zeta\in[\zeta_-,\zeta_+]$, with a reflecting boundary
condition at one side $\partial_\zeta h(\alpha,\zeta_-)=0$ and 
$h(\alpha,\zeta_+)=1$ at the other side (which corresponds actually to 
the absorption at $\zeta=\zeta_+$). The coordinates $\zeta_\pm$ are
the points where the deterministic force and the white noise have equal 
strengths.

Now, we can obtain $h(\alpha,\zeta)$ straightforwardly.
\be
\smean{\EXP{-\alpha\Lambda}} \simeq h(\alpha,\zeta_-) 
= \frac{1}{\cosh\sqrt{\alpha/N(E)}}
\:,\ee
where $N(E)={g}/{2\ln^2(g/k)}$ is the IDoS per unit length.

If $\Lambda\equiv\tilde\Lambda(-\infty)$ is the ``time'' needed to cross
the interval, 
the distance between two nodes of the wave function is a sum of two such
(independent) random variables $\ell=\Lambda_1+\Lambda_2$. Therefore its 
distribution is given by inverse Laplace transformation of 
$\smean{\EXP{-\alpha\ell}}=1/\cosh^2\sqrt{\alpha/N(E)}$.
We obtain~:
\be\label{dol0susy2}
P(\ell) = N(E) \, \varpi_0(N(E) \ell)
\:,\ee
where $\varpi_0(x)$ is the inverse Laplace transform of $\cosh^{-2}\sqrt{s}$~:
\bea\label{pi0susy}
\varpi_0(x) &=&  \heav(x) \sum_{m=0}^\infty
\left[\pi^2(2m+1)^2x - 2\right] \EXP{ -\frac{\pi^2}{4}(2m+1)^2 x } 
\APPROX{x\to\infty} \pi^2 x \EXP{ -\frac{\pi^2}{4} x }\\
&=& \frac{4}{\sqrt{\pi}}\:\frac{\heav(x)}{x^{3/2}} 
\sum_{m=1}^\infty(-1)^{m+1} m^2\EXP{-m^2/x}
\APPROX{x\to0} \frac{4}{\sqrt{\pi}}\:\frac{\heav(x)}{x^{3/2}}\:\EXP{-1/x}
\:,\eea
where $\heav(x)$ is the Heaviside function.

We find the distribution of the ground state energy~:
\be
W_0(E) = L\rho(E)\,\varpi_0(LN(E))
\:,\ee
with the following limiting behaviour~:
\bea \label{W0lim1}
W_0(E) &\simeq& \frac{8}{\sqrt{2\pi gL}} \frac1E 
\exp-\frac{\ln^2(g^2/E)}{2gL} 
\hspace{1.2cm}{\rm for} \hspace{0.5cm} E\ll g^2\EXP{-\sqrt{2gL}}\\
\label{W0lim2}
&\simeq& \frac{8\pi^2g^2L^2}{E\ln^5(g^2/E)}
\exp-\frac{\pi^2gL}{2\ln^2(g^2/E)} 
\hspace{0.5cm} {\rm for} \hspace{0.5cm} g^2\EXP{-\sqrt{2gL}}\ll E\ll g^2
\:.\eea
In ref.~\cite{Tex00}, an integral representation of $W_n(E)$ is also given
together with an explicit expression for $W_1(E)$.

It is also interesting to point out that the ground state distribution is 
characterized by a typical value $E_0^{\typ} \simeq g^2 \EXP{-g L}$, 
a median value $E_0^{\rm med} \sim g^2 \EXP{-\sqrt{gL}}$,
and a mean value
$\smean{E_0} \sim g^{2}\, (gL)^{1/2}\, \EXP{-C(gL)^{1/3}}$
where $C$ is a numerical constant (see ref.~\cite{Tex00}).
This latter expression has also been obtained in ref.~\cite{MonOshComBur96} 
where upper and lower bounds were found using a perturbative 
expression for the ground state energy as a functional of $\phi(x)$.

The distribution (\ref{pi0susy}) was obtained in ref.~\cite{LeDMonFis99} in
the context of the classical diffusion by using a real space renormalization
group method. In this case the distribution is interpreted as the distribution
of the smallest relaxation time.

In summary, these two examples illustrate the fact that extreme value
spectral statistics provides an information on correlations of eigenvalues~:
the extreme value statistics of independent and identically distributed 
random variables have been
classified by Gumbel \cite{Gum54}. Therefore extreme value distribution for
eigenvalues that differ from one of the three Gumbel's laws indicates level
correlations.


\section{Trace formulae, spectral determinant and diffusion on graphs\label{sec:graphs}}

\subsection{Introduction}

Up to now we have discussed several questions related to the physics of 
weak and strong localization. In the previous section we have considered
spectral properties while in the sections \ref{sec:wl} and \ref{sec:expfun}
we mostly discussed transport properties.

In the present section we are going to review several results related to the
study of the Laplace operator on metric graphs.  An object at the core of our
discussion is the spectral determinant of the Laplace operator, formally
defined as $S(\gamma)=\det(\gamma-\Delta)$, where $\gamma$ is a spectral
parameter. This quantity encodes the information on the spectrum of the
Laplace operator on the graph. The subsection \ref{sec:description} recalls
the basic conventions required to describe metric graphs. The subsections
\ref{sec:tf} and \ref{sec:sd} review general results on trace formulae and
spectral determinants. These subsections will appear at first sight quite
technical and unrelated to the previous sections. However we will see that
there exists a close relation with the question of quantum transport~: the
Laplace operator is the generator of the diffusion on the graph and its
spectral properties play a central role in the study of transport. This
connection, already evoked in the introduction and section~\ref{sec:wl}, will
be emphasized again in subsection~\ref{sec:wlrg} where the explicit relation
between quantum transport and spectral determinant is recalled.

One interest for spectral determinants is that they can be used as generating
functions for various quantities characterizing the diffusion on the graph.
This will be illustrated by using the connection with trace formulae
(section~\ref{sec:tf}) and further exploited in subsection~\ref{sec:wlrg} in
the context of quantum transport.  The efficiency of the method stems from the
fact that, even though $S(\gamma)$ seems to be a complicated object at first
sight, involving an infinite number of eigenvalues, it can be expressed as the
determinant of a finite size matrix. This relation, established by Pascaud \&
Montambaux \cite{Pas98,PasMon99}, allows computing easily and systematically
the spectral determinant for arbitrary graphs.

The study of the Laplace operator on metric graphs (or {\it quantum graphs})
is not restricted to transport and
appears in many physical contexts ranging from 
organic molecules \cite{RudSch53}, superconducting networks \cite{Ale83},
phase coherent transport in networks of weakly disordered wires
\cite{DouRam85,PasMon99,AkkComDesMonTex00,TexMon04}, 
transport in mesoscopic networks 
\cite{Sha82,Sha83,ButImrAzb84,GefImrAzb84,AvrSad91,VidMonDou00,TexMon01} 
or metallic agregates \cite{CarAku04}.
We are not going to review this history and refer the interested reader
to the references~\cite{Ale83,AvrRavZur88,Col98,KotSmi99,AkkComDesMonTex00,KotSmi03} 
(for a recent review see the appendix of Pavel Exner in 
ref.~\cite{AlbGesHoeHol04}).  
Mathematical aspects of quantum graphs are discussed in the recent issues of 
Waves Random Media {\bf14} (2004) and Journal of Physics A~{\bf38} (n$^o$22) 
(June 2005). See in particular the refs.~\cite{Kuc04,Kuc05}.

\subsection{Description of metric graphs\label{sec:description}}

\begin{figure}[htbp]
  \centering
  \includegraphics[scale=1.25]{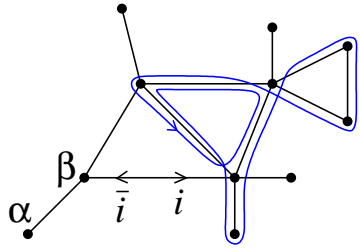}
  \caption{{\it An example of graph with 11 vertices and 13 bonds.
           The arrows  show the orientations of the arc $i$ 
           and the reversed arc $\bar i$.
           The blue curve is an example of (primitive) orbit with one 
           backtracking.}}
  \label{fig:exampleqds}
\end{figure}

We collect some background material from the theory of graphs and establish
our notations and conventions (illustrated on figure~\ref{fig:exampleqds}).

\noindent{\bf Vertices, bonds, arcs.--}
Let us consider a network of $B$ wires (bonds) connected at $V$ vertices. The
latter are labelled with Greek indices $\alpha$, $\beta$,... Therefore, the
bonds of the graph can be denoted by a couple of Greek indices $(\ab)$. The
oriented bonds, denoted as {\it arcs}, will also play an important role. The
two arcs related to the bond $(\ab)$ will be labelled by $\ab$ and $\ba$ or
more simply with Roman letters $i$, $j$,...

\noindent{\bf Adjacency matrix.--}
The basic object is the adjacency
$V\times V$~matrix $a_{\alpha\beta}$ characterizing the topology of the
network~:  $a_{\alpha\beta}=1$ if $\alpha$ and $\beta$ are connected by a
bond~; $a_{\alpha\beta}=0$ otherwise\footnote{
  Note that this definition assumes that two vertices are connected by 
  at most one bond and that a bond never forms a loop.
  We emphasize that {\it this does not imply any particular restriction on 
  the topology of the graph}~: one can always introduce a vertex on 
  a bond, which separates the bond into two bonds, without modifying 
  the properties of the graph. The numbers of vertices and bonds can 
  always be made arbitrary large and this is partly a matter of choice. 
  The case of closed bonds (forming a loop) or vertices connected by 
  multiple bonds requires a simple generalization of the formalism
  presented here (see for example ref.~\cite{Pas98} and appendix C of 
  ref.~\cite{AkkComDesMonTex00}). This generalization allows in 
  particular to minimize $B$ and $V$, which makes the computation 
  sometimes easier.
}.
The connectivity of the vertex $\alpha$, denoted by $m_\alpha$, is 
related to the adjacency matrix by $m_\alpha=\sum_\beta a_{\alpha\beta}$.
Summed over the remaining index, the adjacency matrix gives  the 
number of arcs~: $\sum_{\alpha,\beta} a_{\alpha\beta}=2B$.

\noindent{\bf Orbits.--}
A path is an ordered set of arcs such that the end of an arc coincides with
the begining of the following arc. The equivalence class of all closed paths
equivalent by cyclic permutations is called an {\it orbit}.
An orbit is said to be {\it primitive} when it can not be decomposed as 
a repetition of a shorter orbit.

\noindent{\bf Scalar functions.--}
The graphs we consider here are not simply topological objects but have also
some metric properties~: the bond $(\alpha\beta)$ is characterized by its 
length $l_{\alpha\beta}$ and identified with the interval 
$[0,l_{\alpha\beta}]$ of $\RR$. A scalar function $\psi(x)$ is defined
by  its  components $\varphi_{\ab}(x_{\ab})$ on each bond,
where $x_\ab$ is the coordinate measuring the distance along the bond 
from the vertex $\alpha$ (note that $x_\ab+x_\ba=l_\ab$).

\noindent{\bf Continuous boundary conditions.--}
When studying the Laplace operator $\Delta$ acting on scalar functions, 
boundary conditions at the vertices must be specified in order to 
ensure self adjointness of the operator.
Let us introduce the notations~:
\bea
\label{phiab}
\varphi_\ab &\equiv& \varphi_\ab(x_\ab=0) \\
\label{phiprimeab}
\varphi'_\ab &\equiv& \frac{\D\varphi_\ab}{\D x_\ab}(x_\ab=0)
\:,\eea
for the value of the function and its derivative at the vertex $\alpha$, along
the arc $\ab$. 
The continuous boundary conditions assume~:\\
({\it i}) continuity of the function at each vertex~: all the 
components $\varphi_{\alpha\beta}$ for all $\beta$ neighbours of $\alpha$
are equal. The value of the function at the vertex is denoted 
$\varphi(\alpha)$.\\
({\it ii})
$
\sum_\beta a_{\alpha\beta} \varphi'_{\alpha\beta}=
\lambda_\alpha\varphi(\alpha)
$
where the adjacency matrix in the sum constrains it to run over the
neighbouring vertices of $\alpha$. Therefore the sum runs over all wires
issuing from the vertex. The real parameter $\lambda_\alpha$ allows 
describing several boundary conditions~: $\lambda_\alpha=\infty$ enforces the
function to vanish, $\varphi(\alpha)=0$, and corresponds to the 
Dirichlet boundary
condition. $\lambda_\alpha=0$ corresponds to the Neumann boundary condition.
We refer to the case of a finite $\lambda_\alpha$  as ``mixed boundary 
conditions''.
In the problem of classical diffusion, the Dirichlet condition describes the 
connection to a reservoir that absorbs particles, while the Neumann condition
ensures conservation of the probability current and describes an internal 
vertex.

\noindent{\bf General boundary conditions.--}
Thanks to the continuity hypothesis, the simple boundary conditions we have 
just described allow us to introduce vertex variables $\varphi(\alpha)$ 
(this is convenient since the number of vertices $V$ is  usually smaller than 
the number of arcs $2B$).
However these are not the most general boundary conditions, and in general the
different components associated to the arcs issuing from a given vertex do not
have {\it a priori} the same limit at the vertex. The most general boundary
conditions can be written as~:
\be
C\,\varphi + D\,\varphi' =0
\ee
where $\varphi$ and $\varphi'$ are the column vectors of dimension $2B$
collecting the variables (\ref{phiab}) and (\ref{phiprimeab}), respectively.
$C$ and $D$ are two square $2B\times2B$ matrices. 
In the arc formulation, the information on the topology of the graph is 
encoded in these two matrices.
The self adjointness of the Laplace operator is ensured if these two 
matrices satisfy the following conditions \cite{KosSch99}~:
({\it i}) $CD^\dagger$ is self-adjoint.
({\it ii}) The  $2B\times4B$ matrix $(C,D)$ must have maximal rank.
Note that the choice of $C$ and $D$ is not unique.
The introduction of boundary conditions without continuity at the vertices has 
been motivated physically in 
refs.~\cite{Sha83,GefImrAzb84,ButImrAzb84,AvrExnLas94,Exn95,Exn96a}.

\noindent{\bf Scattering theory interpretation.--}
We can give a more clear physical meaning to these conditions in a
scattering setting. Let us consider the equation 
$-\Delta\varphi(x)=E\varphi(x)$
for a positive energy $E=k^2$. We decompose the component on the bond as the
superposition of an incoming and an outgoing plane wave~:
$\varphi_\ab(x_\ab)=I_\ab\,\EXP{-\I kx_\ab}+O_\ab\,\EXP{\I kx_\ab}$. It is
convenient to collect the amplitudes in column vectors $I$ and $O$. The vector
$I$ contains the incoming plane wave amplitudes and the vector $O$ 
the outgoing plane wave amplitudes. Both are related by a vertex scattering
matrix~:  $O=Q\,I$. The self-adjointness of the Schr\"odinger operator is now
ensured by imposing the unitarity of the scattering matrix~: $Q^\dagger Q=1$.
The relation between the two formulations yields~:
\be\label{relQDC}
Q=(\sqrt{\gamma}\,D-C)^{-1}(\sqrt{\gamma}\,D+C)
\:.\ee 
For the discussion below it is convenient to introduce the parameter
$\gamma$ related to the energy by $E=-\gamma=k^2+\I0^+$ (the matrix $Q$ is
unitary for $E>0$ only). 
An example of a matrix $Q$ for one vertex without continuity of the wave
function was given in refs.~\cite{Sha83,GefImrAzb84,ButImrAzb84}. This
particular choice has become popular in mesoscopic physics (see also 
ref.~\cite{TexMon01} where a more convenient parametrization was provided). 

\noindent
Since we are considering here compact graphs, we are dealing only with discrete
spectrum. The study of noncompact graphs (with some wires of infinite length),
which have continuous spectra, requires a scattering theory approach,
initiated by the work of Shapiro \cite{Sha82,Sha83} and discussed in 
refs.~\cite{GerPav88,AvrSad91,Ada92,KosSch99,KotSmi00,BarGas01,KotSmi03,TexMon01,Tex02,TexBut03,TexDeg03}.

\noindent{\bf Magnetic fluxes.--}
The Laplace operator arises in the context of diffusion equation but also 
in quantum mechanics. In this last case a natural generalization is to 
introduce a magnetic field. This is achieved by introducing a 1-form 
$A(x)\D x$ along the wires (the derivative must then be replaced by
a covariant derivative~: $\D_x\to\Dc_x=\D_x-\I A(x)$). We denote by 
$\theta_\ab=\int_\alpha^\beta\D x\,A(x)$ the corresponding line integral
along the arc $\ab$.
In the context of classical diffusion, magnetic fluxes and winding numbers are
conjugated variables.

\subsection{Trace formulae and zeta functions\label{sec:tf}}

Trace formulae play an important role in spectral theory. A famous example is
the Selberg trace formula which may be viewed as an extension of the Poisson
summation formula to non commutative groups \cite{Sel56}. An analogous formula
in physics is the Gutzwiller trace formula \cite{Gut90} that has been
extensively used in the context of quantum chaos and mesoscopic physics.
Although one is exact and the other only a semiclassical approximation, both
of them express the partition function (or the density of states, whose
Laplace transform gives the partition function) as a sum over closed
geodesics. They provide a connection between quantum properties (spectrum) and
classical properties (classical trajectories). Below we discuss two examples
of exact trace formulae that have been derived for graphs and their relation
to spectral determinants.

\subsubsection*{Roth's trace formula}

It expresses the trace of the heat kernel (partition function
$Z(t)=\tr{\EXP{t\Delta}}$) as an infinite series of contributions of periodic
orbits on the graph. This remarkable formula, due to Roth \cite{Rot83,Rot83a},
applies to graphs with continuous boundary conditions with $\lambda_\alpha=0$.
It it easy to include magnetic fluxes additionally~:
\be\label{Roth}
Z(t)=
\frac{\cal L}{2\sqrt{\pi t}} + \frac{V-B}{2} + \frac{1}{2\sqrt{\pi t}}
\sum_{{\cal C}} l(\widetilde{\cal C}) \alpha({\cal C})
\EXP{-\frac{l({\cal C})^2}{4t}+\I\theta({\cal C})}
\:.\ee
${\cal L}$ is the ``volume'' of the graph, {\it i.e.} the total length
${\cal L}=\sum_{(\ab)}l_\ab$.
The sum runs over all orbits ${\cal C}=(i_1,i_2,\cdots,i_n)$ constructed in
the graph. 
$l({\cal C})=l_{i_1}+\cdots+l_{i_n}$ is the total length of the orbit,
and $\theta({\cal C})$ the magnetic flux enclosed by it.
$\widetilde{{\cal C}}$ designates the primitive orbit associated with a 
given orbit ${\cal C}$.
The weight $\alpha({\cal C})$ depends on the connectivity of the vertices 
visited by the orbit~:
$\alpha({\cal C})=\epsilon_{i_1i_2}\epsilon_{i_2i_3}\cdots\epsilon_{i_ni_1}$.
The matrix $\epsilon$ couples the arcs of the graph\footnote{
  Note that the parameter $\epsilon_{ij}$ has a simple interpretation in 
  the scattering formulation \cite{TexMon01}~: it is the 
  probability amplitude to be transmitted from arc $i$ to arc $j$.
  The arc matrices $\epsilon$ and $Q$ are related by 
  $\epsilon_{ij}=Q_{i\bar j}$ where $\bar j$ denotes the reversed arc.
}~:\\
$\bullet$ if $i$ ends at vertex $\alpha$ and $j$ starts from it, we have 
$\epsilon_{ij}=2/m_\alpha$, where 
$m_\alpha$ is the connectivities of the vertex.\\
$\bullet$ If moreover $i$ and $j=\bar i$ are the reversed arcs
$\epsilon_{i\bar i}=2/m_\alpha-1$.\\
$\bullet$ Otherwise $\epsilon_{ij}=0$.

It is worth mentioning that the Roth trace formula has found recently some
pratical applications to analyze magnetoconductance measurements on large
square networks\footnote{   
  More precisely, the expansion given below by eq.~(\ref{traceformulaS}) for
  the Laplace transform of the partition function is relevant in this case.
}~\cite{FerAngRowGueBouTexMonMai04}.

\subsubsection*{Ihara-Bass trace formula}

Instead of considering metric graphs we now turn to graphs viewed as purely 
combinatorial structures consisting of vertices connected by bonds 
of equal lengths ($l_\ab=1$). 
All the information is therefore encoded in the adjacency matrix $A$.
In this setting, the Ihara $\zeta$-function is defined as
\be
\zeta(u)^{-1}  
= \prod_{\widetilde{\cal C}_B} \left( 1 - u^{l(\widetilde{\cal C}_B)} \right)
\:,\ee
where the infinite product extends over all primitive backtrackless orbits
$\widetilde{\cal C}_B$.
The Ihara-Bass trace formula relates this infinite product to the determinant
of a finite size matrix \cite{Bas92,StaTer96}~:
\be\label{IharaZfct}
\zeta(u)^{-1}  = (1-u^2)^{B-V}\,\det((1-u^2)\ide - u\,A + u^2\,Y)
\ee
where $\ide$ is the identity matrix, 
$A$ is the adjacency matrix ($A_\ab\equiv a_\ab$) and $Y$ is
the diagonal matrix encoding all connectivities~: $Y_\ab=\delta_\ab m_\alpha$.
This relation was derived by Ihara \cite{Iha66} for regular graphs (all
vertices with same connectivity) and later generalized to arbitrary graphs by
Bass~\cite{Bas92}.
The formalism that we have developed for metric graphs is flexible enough
to describe these combinatorial structures in the same setting.
We will see below that the $\zeta$-function is in fact directly related to the 
spectral determinant. Moreover the Ihara-Bass trace formula can be further 
generalized to include backtrackings.

\subsection{Spectral determinant\label{sec:sd}}

In the physics literature spectral determinants arise in evaluating path
integrals which are quadratic in the fluctuation around a given background
field. A well known technique for regularizing such quadratic path integrals
is the $\zeta$-function regularization. Given a certain operator ${\cal O}$
whose eigenvalues $E_n$ are known, one defines the following $\zeta$-function~:
$\zeta(s)=\sum_n E_n^{-s}$. This expression, which converges for $s$
sufficiently large, can be analytically extended to a meromorphic function
regular at the origin. The corresponding regularized determinant is then
$\det{\cal O}=\exp(-\zeta'(0))$ ($\equiv\prod_nE_n$ formally). 
One can find a general
discussion on functional determinants in ref.~\cite{For87}.


In the context of graphs another regularization of the determinant of the
Laplace operator has been used and has proved to be directly related to several
physical quantities. If we introduce the trace of the resolvent
$g(\gamma)=\sum_n(\gamma+E_n)^{-1}$, the spectral determinant is defined as
$S(\gamma)=\exp(\int^\gamma\D\gamma'\,g(\gamma'))$ (it can be formally written
as\footnote{
  The limit $\gamma\to0$ of $S(\gamma)$ has been
  discussed in ref.~\cite{AkkComDesMonTex00}. 
}
$S(\gamma)=\prod_n(\gamma+E_n)$). Moreover, the spectral parameter $\gamma$
has in some cases a physical meaning (see section~\ref{sec:wlrg}).

\mathversion{bold}
\subsubsection*{Laplace operator $\Delta$ with continuous boundary conditions}
\mathversion{normal}

Pascaud \& Montambaux have shown in refs.~\cite{Pas98,PasMon99} that 
$S(\gamma)=\det(\gamma-\Delta)$ can be 
related to the determinant of a $V\times V$-matrix~:
\be\label{spedet}
S(\gamma) =
\gamma^{\frac{V-B}2}
\prod_{(\alpha\beta)}\sinh(\sqrt\gamma l_{\alpha\beta})\:
\det M
\:,\ee
where the product runs over all bonds of the network.
The $V\times V$-matrix $M$ is defined as~:
\be\label{defM}
M_\ab=\delta_\ab
\left(\frac{\lambda_\alpha}{\sqrt\gamma} + \sum_\mu a_{\alpha\mu}
      \coth(\sqrt{\gamma}l_{\alpha\mu})\right)
-a_\ab\frac{\EXP{-\I\theta_\ab}}{\sinh(\sqrt{\gamma}l_\ab)}
\:,\ee
where the adjacency matrix constrains the sum to run over all vertices
$\mu$ connected to $\alpha$. The matrix $M$ encodes all information about the
network~: topology (matrix $a_{\alpha\beta}$), lengths of the wires ($l_\ab$),
magnetic fluxes ($\theta_\ab$), boundary conditions ($\lambda_\alpha$). Below,
we  give several examples which show how $S(\gamma)$ is related to the
characteristic function of various interesting functionals of Brownian
curves on a graph.

The expression (\ref{spedet}) was originally derived in ref.~\cite{PasMon99}
by constructing the Green's function in the graph and eventually integrating
it~:
$\int\D x\,\bra{x}\frac1{\gamma-\Delta}\ket{x}=\drond{}{\gamma}\ln S(\gamma)$.
A more direct derivation using path integral was later obtained  in 
ref.~\cite{AkkComDesMonTex00}.

The relation between the Roth trace formula and the result (\ref{spedet})
was addressed in ref.~\cite{AkkComDesMonTex00}. The main difficulty to
establish this connection is to go from vertex variables to the arc language
of eq.~(\ref{Roth}). A first step is to relate the determinant of the
vertex-matrix, eq.~(\ref{spedet}), to the determinant of an arc-matrix~:
\be\label{LfctS}
S(\gamma)=\gamma^{\frac{V-B}{2}}\EXP{\sqrt{\gamma}\,{\cal L}}
\,\det(\ide - QR)
\:.\ee
$R$ is the $2B\times2B$~matrix 
$R_{ij}=\delta_{i\bar j}\,\EXP{-\sqrt{\gamma} l_i + \I\theta_i}$, where 
$\bar j$ denotes the reversed arc.
The matrix $Q$ was introduced above (eq.~(\ref{relQDC})) and is related to 
$\epsilon$ by $Q_{ij}=\epsilon_{i\bar j}$. 
For the continuous boundary conditions with $\lambda_\alpha=0$ it is given by
$Q_{ii}=2/m_\alpha-1$, 
$Q_{ij}=2/m_\alpha$ if $i$ and $j$ both issue from the vertex $\alpha$.
$Q_{ij}=0$ in other cases.
Eq.~(\ref{LfctS}) holds for the most simple boundary conditions~: 
continuous with $\lambda_\alpha=0$. The general case is discussed below.
Expanding the determinant by using 
$\ln\det(\ide - QR)=- \sum_{n=1}^\infty \frac{1}{n}\tr{(QR)^n}$, 
we eventually express the spectral determinant as an infinite product over the 
primitive orbits~:
\be\label{zetafctS}
S(\gamma)=\gamma^{\frac{V-B}{2}}\EXP{\sqrt{\gamma}\,{\cal L}}
\prod_{\widetilde{\cal C}}
\left(  
  1 - \alpha(\widetilde {\cal C})
  \EXP{-\sqrt{\gamma}\,l(\widetilde {\cal C})+\I\theta(\widetilde {\cal C})}
\right)
\:.\ee
This shows that the spectral determinant is a zeta function 
(references on zeta functions on graphs are refs.~\cite{StaTer96,Che99}).
The last step to connect this formula to Roth's trace formula is to 
notice that 
\be\label{traceformulaS}
\drond{}{\gamma}\ln S(\gamma)=\frac{{\cal L}}{2\sqrt\gamma}+\frac{V-B}{2\gamma}
+\frac{1}{2\sqrt\gamma}\sum_{{\cal C}} l(\widetilde {\cal C})
\alpha({\cal C})\EXP{-\sqrt{\gamma}\,l({\cal C})+\I\theta({\cal C})}
\:,\ee
where the sum now runs over all orbits (if ${\cal C}$ is not primitive, 
$\widetilde{\cal C}$ designates the related primitive orbit).
Finally we perform an inverse Laplace transform of this expression,
$\int_0^\infty\D t\,Z(t)\EXP{-\gamma t}=\drond{}{\gamma}\ln S(\gamma)$
and eventually recover eq.~(\ref{Roth}).
Examples of applications are studied in ref.~\cite{AkkComDesMonTex00}.

\mathversion{bold}
\subsubsection*{Schr\"odinger operator $-\Delta+V(x)$ with general boundary conditions}
\mathversion{normal}

The result (\ref{spedet}) of Pascaud \& Montambaux has been generalized by one
of us. In refs.~\cite{Des00,Des00a} a similar formula was obtained for the
spectral determinant of the Schr\"odinger operator $-\Delta+V(x)$ (the Hill
operator), where $V(x)$ is a scalar potential defined on the graph. In
ref.~\cite{Des01} the formula was further extended to describe general
boundary conditions as well. As an illustration we construct the generating
function of the number of closed orbits with a given number of backtrackings
\cite{Iha66,WuKun99}.

\vspace{0.25cm}

The starting point is to introduce 
two linearly independent solutions of the differential equation
$(-\D_x^2+V_\ab(x)+\gamma)f(x)=0$ on $[0,l_\ab]$. We associate each 
solution with an arc. Let us denote $f_\ab(x_\ab)$ the function
satisfying 
\be
f_\ab(0)=1 
\hspace{0.5cm} \mbox{ and } \hspace{0.5cm}
f_\ab(l_\ab)=0
\:.\ee 
Therefore a second solution of the differential equation is naturally denoted 
$f_\ba(x_\ba)=f_\ba(l_\ab-x_\ab)$.
The Wronskian of these two solutions, defined as
$W_\ab=f_\ab(x_\ab)\frac{\D f_\ba(x_\ba)}{\D x_\ab}
      -\frac{\D f_\ab(x_\ab)}{\D x_\ab}f_\ba(x_\ba)$, 
is constant along the bond~:
$W_\ab=W_\ba=-f'_\ab(l_\ab)=-f'_\ba(l_\ab)$.
If we consider the case $V(x)=0$, the solution is simply
$
f_\ab(x_\ab)
=\frac{\sinh\sqrt\gamma(l_\ab-x_\ab)}{\sinh\sqrt\gamma l_\ab}
\equiv\frac{\sinh\sqrt\gamma x_\ba}{\sinh\sqrt\gamma l_\ab}
$.

All the required information about the potential is contained in the 
$2B\times 2B$ arc-matrix $N$, defined as
\be\label{n161}
N_{\alpha\beta,\mu\eta}  = 
\delta_{\alpha\mu}  \delta_{\beta\eta} 
f'_\ab(0) 
- \delta_{\alpha\eta}\delta_{\beta\mu} f'_\ab(l_\ab)
\:.\ee
This matrix couples a given arc to itself and to its reversed arc, only.
If we assume that {\it the matrices $C$ and $D$ are independent on the 
spectral parameter $\gamma$}, it was shown in ref.~\cite{Des01} that~:
\be\label{n40} 
 S(\gamma ) = \det(\gamma-\Delta+V(x)) = 
 \prod_{(\alpha\beta)}  \frac{1}{\wal}\ \det(C+DN)
\ee
where the product runs over all bonds. 
Functional determinants on a segment of $\RR$ with general boundary conditions
at the boundaries have been studied by McKane \& Tarlie \cite{McKTar95}
using the formalism developed by Forman~\cite{For87}.

It is also interesting to encode the information on the potential
$V(x)$ in the matrix $R$ defined as\footnote{
  Let us remark that, for the free case ($V(x)\equiv 0$),
  we recover the expression of the matrix $R$ given above, coupling the arc 
  $\ab$ to the reversed arc $\ba$ only~:
  $
  R_{\alpha\beta,\mu\eta} 
  =\delta_{\alpha\eta}\delta_{\beta\mu} \EXP{-\sqrt{\gamma} \lab }
  $. 
}~:
\be\label{n41} 
  R \equiv  ( \sqrt{\gamma} \ide + N )( \sqrt{\gamma} \ide - N  )^{-1}
\:.\ee
Then
\bea\label{n42} 
 S(\gamma) =  \prod_{(\alpha\beta)} \frac{1}{ \wal } \:
 \frac{1}{\det( \ide + R )} \: \det (C - \sqrt{\gamma} D ) \:
 \det( \ide - Q R ) 
\eea
where the matrix $Q=(\sqrt{\gamma} D - C )^{-1}(\sqrt{\gamma} D + C )$
was defined above by eq.~(\ref{relQDC}).
Let us explain the structure of eq.~(\ref{n42})~: the term 
$\det( \ide + R )\prod_{(\alpha\beta)}W_\ab$ contains informations on the 
potential only\footnote{
  The matrix $N$ encodes the information about the potential on the bonds 
  through $f'_\ab(0)$ and $f'_\ab(l_\ab)$.
  This information can also be introduced through transmission $t_\ab$ and 
  reflection $r_\ab$ amplitudes by the potential $V_\ab(x)$. This relation
  is developed in ref.~\cite{TexMon01}.
  It is interesting to point out that 
  $\det( \ide + R )\prod_{(\alpha\beta)}W_\ab
  =2^B\gamma^{B/2}\prod_{(\alpha\beta)}t_\ab
  =2^B\gamma^{B/2}\prod_{(\alpha\beta)}R_{\ab,\ba}$.
}.
The factor $\det (C -\sqrt{\gamma}D)$ contains only information on the 
topology of the graph.
The most interesting part is the last term $\det(\ide-QR)$ combining both
informations. In particular this last part generates the infinite
contributions of primitive orbits in (\ref{zetafctS}).

\subsubsection*{Permutation-invariant boundary conditions.} 

In the previous section the matrices describing the graph are $2B\times2B$ arc
matrices. A simplifiaction can be brought by passing to vertex variables. In
the arc formulation, $C$ and $D$ define the topology of the graph~: two arcs
coupled by $C$ and/or $D$ issue from the same vertex. Therefore it is possible
to organize the basis of arcs in such a way that the matrices $C$ and $D$ have
similar block diagonal structures. The matrices are made of $V$ square blocks,
each corresponding to a vertex. A given block, of dimension
$m_\alpha\times{m_\alpha}$ and denoted $C_{\alpha}$ (and $D_{\alpha}$),
corresponds to the $m_\alpha$ arcs issuing from the vertex $\alpha$. If we
assume that the boundary conditions are invariant under any permutation of the
nearest neighbours of $\alpha$, then it is possible to introduce vertex
variables. In this case we can write:
\begin{eqnarray}
   C_{\alpha } &=&  c_{\alpha } \ide  + t_{\alpha }  F_{\alpha }
   \label{bci} \\ 
   D_{\alpha } &=&  d_{\alpha } \ide  + w_{\alpha }  F_{\alpha }
   \label{bci1} 
\end{eqnarray}
where 
$F_{\alpha }$ is a matrix with all
its elements equal to $1$. The boundary conditions at the vertex $\alpha$ are
characterized by the four parameters $c_{\alpha }$, $d_{\alpha }$,
$t_{\alpha }$ and $w_{\alpha }$ (note however that this choice
is not unique).

\vspace{0.25cm}

Now, let us show  that, for boundary conditions
given by eqs.~(\ref{bci},\ref{bci1})
and $V(x)\ne 0$, the spectral determinant can be expressed in terms
of the vertex $V \times V$-matrix.

We proceed as before but, this time, we consider, for each bond, two other
independent solutions, $\cab(x_\ab)$ and $\cba(l_\ab-x_\ab)=\cba(x_\ba)$, of
the equation $(-\D_{x_\ab}^2+V_\ab(x_\ab)+\gamma)\,\chi(x_\ab)=0$ that 
satisfy the following conditions:
\bea
c_{\alpha}\, \chi_\ab(0)     + d_{\alpha}\,\chi'_\ab(0) &=& 1 \\
c_{\beta} \, \chi_\ab(l_\ab) - d_{\beta} \,\chi'_\ab(l_\ab) &=& 0
\label{p4}  
\eea

\noindent
We denote by $\Xi_\ab$ the Wronskian of $\cab$ and $\cba$.
Following the same steps as before, we  get 
the spectral determinant (up to a multiplicative constant)~:
\be\label{p19} 
 S(\gamma) =  \prod_{(\alpha\beta)}\frac{1}{ \Xi_\ab }\ \det M
\ee
where $M$ is the $V \times V$-matrix~:
\begin{eqnarray} \label{p12}
  M_{\ab} = 
  \delta_{\ab}
  \left(
    1 + \sum_\mu a_{\alpha\mu} 
        \left[
          t_{\alpha }\,\chi_{\alpha\mu}(0)+w_{\alpha }\,\chi_{\alpha\mu}'(0)
        \right]
  \right)
  + a_\ab\, 
  \left[ c_{\alpha } w_{\alpha } -  t_{\alpha } d_{\alpha } \right]\,\Xi_\ab 
\:.
\end{eqnarray}
The eq.~(\ref{n40}), expressing the determinant in terms of arc matrices, and
the eq.~(\ref{p19}), expressing it in terms of vertex matrix, have been
derived up to multiplicative constants independent on $\gamma$. We can
establish a precise relation by comparing their behaviour for
$\gamma\to\infty$. We end up with~:
\begin{equation}\label{p20} 
 \prod_{(\alpha\beta)}\frac{1}{ \wal }\ \det(C+DN)
 =  \prod_{(\alpha\beta)}\frac{1}{ \Xi_\ab }\ \det M
\:.
\end{equation}

\mathversion{bold}
\subsubsection*{Free case ($V(x)=0$). Applications~: counting backtrackings}
\mathversion{normal}

In this subsection we show an application of generalized boundary conditions
to count backtrackings (the figure~\ref{fig:exampleqds} shows an 
orbit with one backtracking).

We now consider  the case $V(x)\equiv0$ still with 
 permutation-invariant boundary conditions.

\noindent
With the notations
$$
    \eta_{\alpha } = \frac{c_{\alpha }   +  \sg d_{\alpha } }
                            { c_{\alpha }  -  \sg d_{\alpha } }
         \qquad , \qquad                      
     \rho_{\alpha } = \frac{\mu_{\alpha }^- - \mu_{\alpha }^+ }
                            {1+ m_{\alpha } \mu_{\alpha }^- }
    \qquad \mbox{and} \qquad
\mu_{\alpha }^{\pm } = \frac{t_{\alpha }   \pm  \sg w_{\alpha } }
      { c_{\alpha }  \pm  \sg d_{\alpha } }
$$
eqs.~(\ref{p20}, \ref{n41}, \ref{relQDC}) lead to~:
\be\label{p804} 
\det (\ide - Q R) = 2^{-V} \prod_{\alpha } \left( \rho_{\alpha }
\eta_{\alpha } \right) \prod_{(\alpha\beta)}\left( 
 1- \eta_{\alpha } \eta_{\beta } \EXP{- 2 \sg \lab}   \right)  \; 
 \det{M} 
\ee
and the $V\times V$-matrix $M$ takes the form~:
\bea
\label{p805}
{M}_{\alpha\beta} = 
\delta_{\alpha\beta} \left(
  \frac{2}{ \rho_{\alpha }\eta_{\alpha } } - 
  \frac{ m_{\alpha }}{\eta_{\alpha } } +  \frac{1}{ \eta_{\alpha }} 
  \sum_\mu  a_{\alpha\mu}
  \frac{1+\eta_{\alpha }\eta_{\mu} \EXP{-2\sg l_{\alpha\mu}}  }
       {1-\eta_{\alpha }\eta_{\mu } \EXP{-2\sg l_{\alpha\mu}} }
\right)
-a_{\alpha\beta}
   \frac{2\,  \EXP{- \sg l_{\alpha\beta }}                         }
        {1-\eta_{\alpha }\eta_{\beta }\EXP{-2\sg l_{\alpha\beta }} }
\:.\eea
Note that the expression (\ref{defM}) is recovered 
for $\eta_\alpha=1$ and $\rho_\alpha=2/(m_\alpha+\lambda_\alpha/\sqrt\gamma)$.

For permutation-invariant boundary conditions, the matrices $C$, $D$ and
 $Q$  (eq.~(\ref{relQDC})) are block-diagonal. The block $Q_{\alpha }$ 
takes the simple form:
\be\label{p900}
     Q_{\alpha } = \eta_{\alpha } 
     \left( -\ide + \rho_{\alpha }  F_{\alpha } \right)
\ee
The only non-vanishing elements of the matrix $QR$ are~:
\be\label{p901}
    (QR)_{\alpha\beta,\mu\alpha}= \left(
    \rho_{\alpha }\eta_{\alpha } - \eta_{\alpha } \; \delta_{\beta\mu } 
 \right) \; \EXP{- \sg l_{\alpha\mu }} 
\ee

\vskip.3cm

\noindent
We call 
$\rho_{\alpha }\eta_{\alpha }$  the transmission factor
at vertex $\alpha $ and $\rho_{\alpha}\eta_{\alpha}-\eta_{\alpha}$
the reflection factor.
By using the same expansion as above we write
\be\label{p902}
 \det(\ide - QR ) 
= \prod_{\widetilde{\cal C}} \left( 1- \mu(\widetilde{\cal C} )
 \EXP{-\sqrt\gamma l(\widetilde{\cal C})} \right)
\:,\ee
where the product is taken over all primitive orbits $\widetilde{C}$ whose
lengths are denoted by $l(\widetilde{C})$.
An orbit being a succession of arcs
$\ldots  ,\tau\alpha,\alpha\beta,\ldots $  with, in $\alpha$,
a reflection (if $\tau =\beta $) or a transmission  (if $\tau \ne \beta $),
the weight $\mu(\widetilde{C})$, in eq.~(\ref{p902}),
will be the product of
all the reflection -- or transmission -- factors along~$\widetilde{C}$.

\vspace{0.25cm}

\noindent 
{\it Graphs with wires of equal lengths.--}
We consider the case of equal lengths $\lab=l$~; then we can choose
$\gamma=1$ without loss of generality and introduce the notation
$u\equiv\EXP{-l}$.
It is clear from eq.~(\ref{p900}) that a backtracking at vertex $\alpha$ 
brings a factor $\eta_{\alpha}$ in the weight $\mu(\widetilde{C})$. 
In order to count backtrackings one has to choose the boundary conditions
$\rho_{\alpha }\eta_{\alpha }=1$ and $\eta_{\alpha}=\eta$. Eq.~(\ref{p804})
takes the simple form~:
\be\label{p903}
\prod_{\widetilde{C}_m} \left( 1 - (1-\eta )^{n_R(\widetilde{C}_m )} u^m
 \right) = (1- \eta^2 u^2 )^{B-V} \det 
 \left( ( 1- \eta^2 u^2  ) \ide +\eta u^2\, Y - u\, A   \right)
 \equiv  Z^{-1}
\ee
$m$ is the number of arcs of the primitive orbit $\widetilde{C}_m$ and
$n_R(\widetilde{C}_m)$ is the number of reflections (backtrackings) 
occuring along  $\widetilde{C}_m$. 
$Y$ is the $V\times V$-matrix 
$Y_{\alpha\beta }=\delta_{\alpha\beta } \; m_{\alpha }$ and $A$ is the
adjacency matrix.

Setting $\eta=1$ implies $n_R(\widetilde{C}_m)=0$ in the left-hand side of
eq.~(\ref{p903})~: we recover Ihara-Bass formula \cite{Iha66,Bas92,StaTer96} 
where
only primitive orbits without tails and backtrackings are kept. (Ihara
\cite{Iha66} established this formula for a regular graph~; the proof for a
general graph is given in refs.~\cite{Bas92,StaTer96} using a direct counting
technique).

 
\vspace{0.25cm}

Now, let us consider closed random walks with a given number
of backtrackings.

Eq.~(\ref{p903}) provides a non-trivial generalization of the Ihara-Bass
formula (an independent derivation is also given in ref.~\cite{Bar99}).
As an application let us consider the problem of enumerating $m$-steps
random walks with $p$-backtracking steps~\cite{WuKun99}.
Taking $Z$ in (\ref{p903}), we get: 
\be\label{p1949}
u \; \frac{ \D \ln Z}{ \D u} = \sum_{m=2}^{\infty } \; \sum_{p=0}^m \; 
\sum_{\alpha =1}^V \; N_m^p(\alpha ) \; (1-\eta )^p \; u^m 
\ee
\noindent 
where $ N_m^p(\alpha ) $ is the number of $m$-steps closed 
random walks on the graph starting at $\alpha$, with $p$ backtrackings.

\vspace{0.25cm}

\begin{figure}[!ht]
\begin{center}
\includegraphics[scale=0.6]{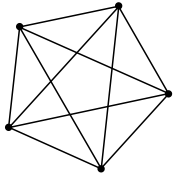}
\end{center}
\caption{\it The complete graph $K_5$ (the $4$-simplex)~: 
         all vertices are connected to each 
         other by bonds of equal lengths.\label{fig:k5}}
\end{figure}

\noindent 
{\it Example~: The complete graph.--}
For the complete graph\footnote{
  The complete graph $K_V$ with $V$ vertices is the $V-1$-simplex~: 
  each vertex is connected to all other vertices (see figure~\ref{fig:k5}).
} $K_V$, we get the results: 
\bea
  N_2^0(\alpha) &=& 0                              \nonumber \\
  N_3^0(\alpha) &=&    (V-1)(V-2)                  \nonumber \\
  N_4^0(\alpha) &=&    (V-1)(V-2)(V-3)             \nonumber \\
  N_5^0(\alpha) &=&    (V-1)(V-2)(V-3)(V-4)        \nonumber \\
  N_6^0(\alpha) &=&    (V-1)(V-2)(V^3-9V^2+29V-32) \label{p1950} 
\eea
\noindent 
and also:
\bea
 N_2^1(\alpha) &=&  N_3^1(\alpha ) \ = \ N_4^1(\alpha ) \ = \ 0   \nonumber \\
 N_5^1(\alpha) &=&  5(V-1)(V-2)(V-3)     \nonumber \\
 N_6^1(\alpha) &=&  6(V-1)(V-2)(V-3)^2  \label{p1952} 
\eea
Note that these expressions have been obtained when vertices are all 
characterized by the same parameter $\eta$. Introducing different parameters
$\eta_\alpha$ allows counting the backtrackings at a given vertex.

\subsection{Quantum chaos on metric graphs}

It has been recently realized that metric graphs are interesting 
models for  quantum chaos. The paper of Kottos \& Smilansky
\cite{KotSmi97} has stimulated several works on spectral statistics and
level correlations \cite{KotSmi99,BarGas00,Tan01b}, and  
renewed the interest in the Roth trace formula \cite{Rot83}.
Progress in the understanding of universality of spectral statistics for
generic quantum graphs has been achieved by Gnutzmann \& Altland
\cite{GnuAlt04}. One of the input of this work is the observation that the
spectral average for a given graph with incommensurate bond lengths is
equivalent to an average over a certain ensemble of unitarity matrices. Star
graphs \cite{BerKea99,KeaMarWin03} have provided simple examples of systems
with intermediate level statistics similar to the one observed in {\v S}eba
billiards and a precise connection has been established in
ref.~\cite{BerBogKea01}. There is however an interesting class of graphs which
do not enter in this category \cite{Tan05}. Chaotic scattering and transport
properties in open graphs (with some infinitely long wires) have also been
studied in refs.~\cite{KotSmi00,BarGas01,KotSmi03}. Since 
quantum chaos is not the
central subject of our review,  this list of papers is not exhaustive~: we
refer the reader interested in this topic to the review
papers~\cite{KotSmi99,KotSmi03,Tan05} and the PhD thesis~\cite{Win03}.

\subsection{Weak localization and Brownian motion on graphs\label{sec:wlrg}}

Our initial physical motivation for the study of the spectral determinant on
graphs was based on the observation that the weak localization correction to
the conductivity is directly expressed in terms of the spectral determinant.
We see from eq.~(\ref{WL}) that
\be\label{WLsd} 
\smean{\Delta\sigma}
= - \frac{2 e^2}{\pi\,{\rm Volume}}
\drond{}{\gamma} \ln S(\gamma)   
\:,\ee
where the spectral parameter is related to the phase coherence
length $\gamma=1/L_\varphi^2$.
The expressions (\ref{WLsd},\ref{spedet},\ref{defM}), due to Pascaud \&
Montambaux \cite{Pas98,PasMon99,AkkComDesMonTex00}, improve the
approach initiated by Dou{\c c}ot \& Rammal \cite{DouRam85,DouRam86}.

\subsubsection*{Nonlocality of the quantum transport in arbitrary networks}

We must stress that the formula (\ref{WLsd}) corresponds to a uniform
integration of the cooperon, defined as 
$P_c(x,x)=\bra{x}\frac1{\gamma-\Delta}\ket{x}$, on
the graph $\smean{\Delta\sigma}\propto-\int\D x\,P_c(x,x)$. This approach is
limited to the case of regular graphs, where all wires play the same role. In
other terms, for a nonregular network, the quantity (\ref{WLsd}) does not
correspond to a quantity measured in a transport experiment. For arbitrary
networks, the cooperon must be integrated over the network with appropriate
nontrivial weights that depend on the topology of the whole network and the
way it is connected to external contacts. Such a generalization was provided in
ref.~\cite{TexMon04}.

\subsubsection*{Windings in a loop connected to a network} 

As we have mentioned in the section~\ref{sec:brownian2},  magnetoconductance 
oscillations due to a magnetic flux and winding properties are
closely related.
For example, if we consider an isolated ring of perimeter $L$ pierced by a 
magnetic flux $\phi$, the well-known behaviour of the harmonics
\begin{equation}
  \label{eq:AAS}
  \smean{\Delta\sigma_n}=\int_0^{2\pi}\frac{\D\theta}{2\pi}
  \smean{\Delta\sigma(\theta)}\,\EXP{-\I n\theta}
  \propto\EXP{-|n|L/L_\varphi}
\end{equation} 
is a direct consequence of the fact that the winding around the ring scales 
with time as $n_t\propto t^{1/2}$ (normal diffusion).
$\theta=4\pi\phi/\phi_0$ is the reduced flux  and $\gamma=1/L_\varphi^2$.
This effect was predicted by Al'tshuler, Aronov \& Spivak (AAS) in 
ref.~\cite{AltAroSpi81} and observed in
experiments on cylinder films~\cite{ShaSha82,AltAroSpiShaSha82}.

Recently it has been noticed that the fact that the ring is connected to arms,
which is necessary to perform a transport experiment, can strongly affect the
harmonics. If we consider a ring connected to $N_a$ long arms, in the limit
$L_\varphi\ll L$, the harmonics are still given by the AAS behaviour
(\ref{eq:AAS}), however when the perimeter is smaller than $L_\varphi$, the
harmonics behave as
$\smean{\Delta\sigma_n}\propto\EXP{-|n|\sqrt{N_a L/L_\varphi}}$.
To clarify the origin of this behaviour, the winding of Brownian trajectories
around a ring connected to another network has been recently examined in
ref.~\cite{TexMon05}. This analysis is based on the fact that the winding
number distribution can be expressed in terms of the spectral determinant. Let
us briefly describe this approach.

\begin{figure}[!ht]
\begin{center}
\includegraphics[scale=1]{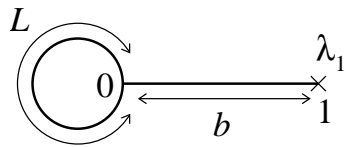}
\end{center}
\caption{\it A diffusive ring attached to a long arm. The parameter $\lambda_1$
         describes the boundary condition at vertex 1~: $\lambda_1=\infty$
         for the Dirichlet boundary and $\lambda_1=0$ for the Neumann boundary
         condition.
         \label{fig:ringwitharm}}
\end{figure}

\begin{figure}[!ht]
\begin{center}
\includegraphics[scale=0.5]{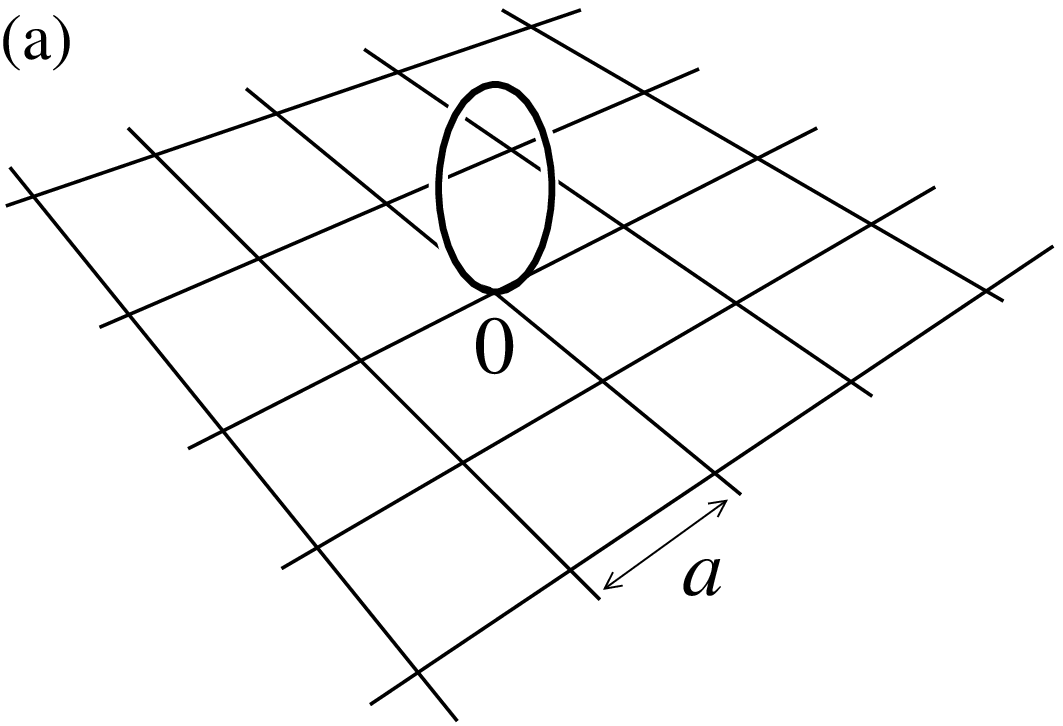}
\hspace{1cm}
\includegraphics[scale=0.5]{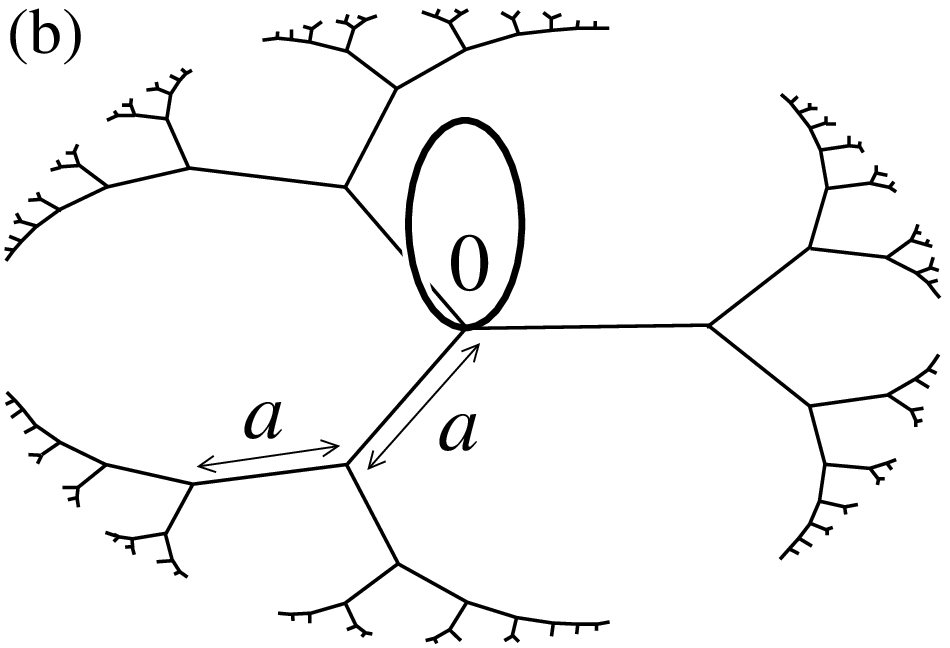}
\end{center}
\caption{\it A diffusive ring attached to {\rm (a)~:} an infinite square 
         lattice,
         {\rm (b)~:} an infinite Bethe lattice of connectivity $z=3$. 
         To ensure that the Bethe lattice is regular, it must be embedded
         in a constant negative curvature surface.
         \label{fig:ringmesh}}
\end{figure}

We consider a ring to which is attached an arbitrary network at vertex $0$
(the figures \ref{fig:ringwitharm} and \ref{fig:ringmesh} give examples
of such a situation). Our aim is to understand how the winding around the ring
is affected by the presence of the network. To answer this question we
introduce the probability to start from a point $x$ on the graph and come back
to it after a time $t$, conditioned to wind $n$ times around the loop~:
\be 
{\cal P}_n(x,t|x,0) 
= \int_{x(0)=x}^{x(t)=x}{\cal D}x(\tau)
\:\EXP{-\frac14\int_0^t\D\tau\,\dot x(\tau)^2} 
\delta_{n,{\cal N}[x(\tau)]} 
\ee
where ${\cal N}[x(\tau)]$ is the winding number of the path
$x(\tau)$ around the loop.  For simplicity, we consider the case where the
initial point is the vertex $0$.
The computation of ${\cal P}_n(0,t|0,0)$ requires some local information
(the eigenfunctions of the Laplace operator at point $x=0$). 
On the other hand the spectral determinant encodes a global information
since it results from a spatial integration of the Green's function
of the Laplace operator over the network. 
However we have shown in ref.~\cite{TexMon05} that if we consider mixed 
boundary conditions at vertex $0$ described by a parameter $\lambda_0$
(see the definition of the boundary conditions given above), 
we can extract the probability of interest\footnote{
  The fact that some ``local'' information, such as ${\cal P}_n(0,t|0,0)$,
  can be extracted from a more ``global'' object, like the spectral 
  determinant, by introducing the mixed boundary conditions with parameter
  $\lambda_0$ has been also used in the context of scattering theory in
  graphs in refs.~\cite{TexBut03,TexDeg03}. 
  The derivative in eq.~(\ref{P0S}) can be understood as a functional
  derivative since $\lambda_0$ plays a role similar to the weight of 
  a $\delta$-potential at $x_0$. 
  The use of functional derivatives in scattering theory for 
  mesoscopic systems has been fruitfully used by
  B\"uttiker and co-workers (see ref.~\cite{But02} and references therein).
} as follows~:
\be\label{P0S}
\int_0^\infty\D t\,{\cal P}_n(0,t|0,0)\,\EXP{-\gamma t}
=\int_{0}^{2\pi}\frac{\D\theta}{2\pi}\:\EXP{-\I n\theta}\:
\frac{\D}{\D\lambda_0}\ln S^{(\lambda_0)}(\gamma)\Big|_{\lambda_0=0}
\:,\ee
where $S^{(\lambda_0)}(\gamma)$ is the spectral determinant for 
mixed boundary condition at vertex $0$ and with a magnetic flux 
$\theta$ piercing the ring.
This formula allows us to express the probability as
\be\label{Wn00}
  \int_0^\infty\D t\:{\cal P}_n(0,t|0,0)\:\EXP{-\gamma t} = 
  \frac{1}{2\sqrt\gamma}
  \frac{\sinh\sqrt\gamma L}{\sinh\sqrt{\gamma} L_{\rm eff}(\gamma)}
  \:\EXP{-|n|\sqrt{\gamma} L_{\rm eff}(\gamma)} 
\:,\ee
where all the information about the nature of the network attached to the 
ring is contained in the effective perimeter $L_{\rm eff}(\gamma)$, defined 
as~:
\be\label{leffnet}
\cosh\sqrt{\gamma} L_{\rm eff}(\gamma)
=\cosh\sqrt{\gamma}L
+\frac{\sinh\sqrt{\gamma}L}{2\,({M}_{\rm net}^{-1})_{00}}
\:.\ee
The matrix $M_{\rm net}$ describes the network in the absence of the loop
(it is given by eq.~(\ref{defM})~; for the network of 
figure~\ref{fig:ringmesh}a, ${M}_{\rm net}$ describes the infinite 
square network without the ring).
The matrix element $({M}_{\rm net}^{-1})_{00}$ has a clear
meaning~: it is the Green function of the Laplace operator in the 
network (without the loop) computed  at the position where 
the ring is attached~:
$\sqrt{\gamma}(M_{\rm net}^{-1})_{00}=\bra{0}\frac1{\gamma-\Delta}\ket{0}$. 

The effective perimeter $L_{\rm eff}(\gamma)$ probes the winding at time
scale $t\sim1/\gamma$~: precisely, the winding number scales with time as
$n_t\sim\sqrt{t}/L_{\rm eff}(1/t)$.

\vspace{0.25cm}

Two interesting examples are

\vspace{0.25cm}

\noindent$\bullet$ {\it Ring attached to an infinite wire 
(figure~\ref{fig:ringwitharm}).--}
When a wire of length $b$ with the Dirichlet boundary at one end is 
attached to the ring, it is easy to see that eq.~(\ref{defM}) gives
$({M}_{\rm net}^{-1})_{00}=\tanh\sqrt\gamma b\leadto{b\to\infty}1$.
At large time $t\gg L^2$, the effective perimeter behaves 
as $L_{\rm eff}\simeq\sqrt{L}\gamma^{-1/4}$. This behaviour is related 
to a scaling of the winding number with  time
\be\label{scalingwire}
n_t \propto t^{1/4}
\:.\ee
The full distribution for the winding number $n$ is given in
ref.~\cite{TexMon05}. The exponent $1/4$  that characterizes anomalously slow
winding around the loop originates from the fact that the diffusive trajectory
spends a long time in the infinite wire, which increases the effective
perimeter at such time scales. This problem is also related to the anomalous
diffusion along the skeleton of a comb, studied in
refs.~\cite{WeiHav86,BalHavWei87} by different methods. It is interesting to
use this picture. Let us consider a random walk along the sites of a line where
the diffusive particle is trapped during a time $\tau$ on each site. The
trapping time is distributed according to a broad distribution
$P_1(\tau)\propto\tau^{-1-\mu}$ with $0<\mu<1$. It follows that the distance
scales with time as $n_t\sim t^{\mu/2}$ \cite{BouGeo90}. If we go back to the
problem of diffusion along the skeleton of a comb (or the winding in the ring
connected to the long arm), the arm plays the role of the trap. The
distribution of the trapping time is given by the first return probability of
the one-dimensional diffusion~:  $P_1(\tau)\propto\tau^{-3/2}$ and we recover
eq.~(\ref{scalingwire}).

\vspace{0.25cm}

\noindent$\bullet$ {\it Ring attached to a square network
(figure~\ref{fig:ringmesh}a).--}
When studying the winding around the loop, it is important to know
whether the Brownian motion inside the network attached to the ring is
recurrent or not. Let us consider the case where the network attached is a
$d$-dimensional hypercubic network. For $d>2$ the Brownian motion is
known to be transient where as, for $d=2$ it is neighbourhood recurrent
(in $d=1$ the Brownian motion is pointwise recurrent). Therefore we
expect the dimension 2 to play a special role. In the large time limit, when 
$t\gg{L^2}$ and $t\gg{a^2}$ ($a$ is the lattice spacing), we find an 
effective length
$\sqrt\gamma\, L_{\rm eff}\simeq\sqrt{\frac{2\pi L}{a\ln(4/\sqrt{\gamma}a)}}$,
that corresponds to a scaling of the winding around the loop
\be\label{scalingsn}
n_t \propto (\ln t)^{1/2}
\:.\ee
This result can be obtained in the same way as for the ring connected to 
the arm. This time the plane acts as a trap. 
The distribution of the trapping time is given by the first return
probability on a square lattice, which is known to behave at large times 
like $P_1(\tau)\propto1/(\tau\ln^2\tau)$ \cite{BarTac93}, from which
we can recover eq.~(\ref{scalingsn}).

\subsubsection*{Occupation time and local time distribution on a graph} 

Another set of problems concern occupation times~: {\it i.e.} time spent by a
Brownian particle in a given region. An example of such a problem is provided
by the famous arc-sine law for the 1d Brownian motion that gives the
distribution of the time spent by a Brownian motion
$(x(\tau),0\leq\tau\leq{t}\,|\,x(0)=0)$ on the half line $\RR^+$. This result
was derived long ago by P.~L\'evy \cite{Lev48}. It has been extended by Barlow,
Pitman \& Yor \cite{BarPitYor89} for a particular graph (star graph with arms
of infinite lengths)~: instead of an infinite line, these authors consider $n$
semi-infinite lines originating from the same point and study the joint
distribution of the times spent on each branch. More recently, this problem
has been reconsidered in the case of arbitrary graphs \cite{Des02}. It may be
stated as follows~: consider a Brownian motion $x(\tau)$ on a graph, starting
from a point $x_0$ at time $0$ and arriving at a point $x_1$ at time $t$. Let
$T_{\ab}$ denotes the time spent on the wire $(\ab)$. This functional is
defined as $T_{\ab}[x(\tau)]=\int_0^t\D\tau\,{\rm Y}_{\ab}(x(\tau))$ where the
function ${\rm Y}_{\ab}(x)$ is $1$ for $x\in(\ab)$ and $0$ otherwise. Our aim
is to compute the Laplace transforms of the joint distribution
\begin{equation}
  \label{eq:octimedef}
  \mean{ \EXP{-\sum_{(\ab)} \xi_\ab\,T_{\ab}} }
  = \int\D x_1\, {\cal F}(x_1,x_0;t;\{\xi_\ab\}) 
\end{equation}
with
\begin{equation}
  {\cal F}(x_1,x_0;t;\{\xi_\ab\})
  =\int_{x(0)=x_0}^{x(t)=x_1} {\cal D}x(\tau)\,
  \EXP{  -\frac14\int_0^t\D\tau\,\dot x^2
         -\sum_{(\ab)} \xi_\ab\,T_{\ab}[x(\tau)]
      }
\:.\end{equation}
The definition (\ref{eq:octimedef}) takes into account averaging over the
final point $x_1$. A conjugate parameter $\xi_\ab$ is introduced for each
variable $T_{\ab}$, {\it i.e.} each bond. A closed expression of the Laplace
transform of the joint distribution (\ref{eq:octimedef}) has been derived in
ref.~\cite{Des02}. The result is given as a ratio of two determinants,
an expression reminiscent of the one of Leuridan \cite{Leu00}, although the 
connection is not completely clear.

Similar methods have been also applied in ref.~\cite{ComDesMaj02} to study the
distribution of the local time
$T_{x_0}[x(\tau)]=\int_0^t\D\tau\,\delta(x(\tau)-x_0)$.

\vspace{0.25cm}

A simplification occurs when the final point coincides with the initial point.
Then it is possible to relate the characteristic function to a single spectral
determinant. We do not develop the general theory here but instead consider an
example close to the one studied in the previous subsection.  Let us consider
the graph in figure~\ref{fig:ringwitharm} and ask the following question~: for
a Brownian motion starting from $x_0$ at time $0$ and coming back to it at 
time $t$,
what is the distribution of the time $T_{\rm arm}[x(\tau)]$ spent in the arm
if in addition the Brownian motion is constrained to turn $n$ times around the
ring~? It is natural to introduce the following function~:
\begin{equation}
  \label{eq:winoc}
  {\cal F}_n(x_0,x_0;t,\xi) =
  \int_{x(0)=x_0}^{x(t)=x_0} {\cal D}x(\tau)\,
  \EXP{  -\int_0^t\D\tau\,
        \left( 
          \frac14\dot x^2 + \xi\,{\rm Y}_{\rm arm}(x)
        \right)
      }\:
  \delta_{n,{\cal N}[x(\tau)]} 
\end{equation}
where ${\cal N}[x(\tau)]$ is the winding number around the ring.
This function is related to the Laplace transform of the distribution
of the functional $T_{\rm arm}[x(\tau)]$ 
\begin{equation}
  \label{disTbras}
  \mean{\EXP{-\xi\, T_{\rm arm}[x]}}_{{\cal C}_n} 
  = \frac{{\cal F}_n(x_0,x_0;t,\xi)}{{\cal F}_n(x_0,x_0;t,0)}
\:,\end{equation}
where $\mean{\cdots}_{{\cal C}_n}$ denotes averaging over curves of 
winding~$n$. The denominator ensures normalization.
The Laplace transform of (\ref{eq:winoc}) is given by a relation similar
to eq.~(\ref{P0S})
\begin{equation}
  \int_0^\infty\D t\,{\cal F}_n(x_0,x_0;t,\xi)\,\EXP{-\gamma t}
  =\int_{0}^{2\pi}\frac{\D\theta}{2\pi}\:\EXP{-\I n\theta}\:
  \frac{\D}{\D\lambda_0}\ln S^{(\lambda_0)}(\gamma)\Big|_{\lambda_0=0}  
\end{equation}
where the appropriate spectral determinant is built as follows~:
({\it i}) since the starting point is fixed at $0$, we introduce mixed 
boundary conditions at this point, with a parameter $\lambda_0$ that will be
used to extract the ``local information''.
({\it ii}) A magnetic flux $\theta$ is introduced (conjugate to the winding
number).
({\it iii}) The spectral parameter is shifted in the arm as 
$\gamma\to\gamma+\xi$ to introduce the variable $\xi$ conjugate to the time 
$T_{\rm arm}$.

We choose the vertex $0$ as initial condition and impose the Dirichlet
boundary condition (which is achieved by setting $\lambda_1=\infty$) at the
end of the arm of length $b$. The spectral determinant is found
straightforwardly (for an efficient calculation of $S(\gamma)$ for a graph
with loops, see ref.~\cite{Pas98} or appendix~C of
ref.~\cite{AkkComDesMonTex00})~:
\begin{equation}
  S^{(\lambda_0)}(\gamma) =
  \frac{\sinh\sqrt{\gamma} L\,\sinh\sqrt{\gamma+\xi}\,b}
       {\sqrt{\gamma}\sqrt{\gamma+\xi}}
  \left[
    2\sqrt{\gamma}\frac{\cosh\sqrt{\gamma} L-\cos\theta}{\sinh\sqrt{\gamma} L}
    +\lambda_0
    +\sqrt{\gamma+\xi}\coth\sqrt{\gamma+\xi}\,b
  \right]
\end{equation}
The terms in the brackets correspond to the matrix\footnote{
  The introduction of the conjugate parameters $\{\xi_\ab\}$ corresponds to
  shift the spectral parameter $\gamma$ on each wire as
  $\gamma\to\gamma+\xi_\ab$.  The matrix to be generalized is not $M$, given
  by eq.~(\ref{defM}), but ${\cal M}=\sqrt{\gamma}M$.
} ${\cal M}$ (since $\lambda_1=\infty$ we 
can consider only the element ${\cal M}_{00}$). The first term is the 
contribution of the loop, the second comes from the boundary condition and 
the last one comes from the arm. It immediately follows that 
\begin{equation}
  \label{relation1}
  \int_0^\infty\D t\,{\cal F}_n(0,0;t,\xi)\,\EXP{-\gamma t}
  =  
  \frac{1}{2\sqrt\gamma}
  \frac{\sinh\sqrt\gamma L}{\sinh\sqrt{\gamma} L_{\rm eff}}
  \:\EXP{-|n|\sqrt{\gamma} L_{\rm eff}} 
\end{equation}
with
\begin{equation}
  \label{Leffwinoc}
  \cosh\sqrt\gamma L_{\rm eff} = \cosh\sqrt\gamma L
  +\frac12\sqrt{1+{\xi}/{\gamma}}\,\sinh\sqrt{\gamma} L\,
   \coth\sqrt{\gamma+\xi}\,b
  \:.
\end{equation}
Let us consider an infinitely long arm $b\to\infty$. If we are interested on
time scales $t\gg\tau_L$, where $\tau_L=L^2$ is the Thouless time over which
the ring is explored, eq.~(\ref{Leffwinoc}) gives 
$\sqrt{\gamma}L_{\rm eff}\simeq(\gamma+\xi)^{1/4}\sqrt{L}$, therefore from
eq.~(\ref{relation1}) we see that 
${\cal F}_n(0,0;t,\xi)\simeq\EXP{-\xi t}\,{\cal F}_n(0,0;t,0)$.
The inverse Laplace transform of eq.~(\ref{disTbras}) leads to
$\smean{\delta(T-T_{\rm arm}[x])}_{{\cal C}_n}\simeq\delta(T-t)$, which means
that the Brownian motion spends almost all the time in the arm (this simple
result confirms the picture presented to explain the scaling of the winding
around the ring of the form $n_t\propto t^{1/4}$).


\section{Planar Brownian motion and charged particle in random magnetic field\label{sec:magimp}}

A model of random magnetic field describing a charged particle moving in a
plane and subjected to the random magnetic field 
${\cal B}(\vec r)=\vec\nabla\times \vec A(\vec r)$ due to an ensemble of
magnetic Aharonov-Bohm vortices is described by the following
Hamiltonian~\cite{DesFurOuv95,Fur97}~:
\bea\label{Him}
H   = \frac12\left(\vec p - \vec A(\vec r) \right)^2 
    + \frac12 \, {\cal B}(\vec r) 
    = \frac12
\left(
  \vec p - \alpha\sum_i 
  \frac{\vec u_z\times(\vec r-\vec r_i)}{(\vec r-\vec r_i)^2}
\right)^2 
+\pi\alpha\sum_i \delta(\vec r-\vec r_i)
\eea
where $0\leq\alpha<1$. We have set $\hbar=e=m=1$.
$\alpha$ is the magnetic flux per vortex, in unit of the quantum flux 
$\phi_0=h/e$~:  $\alpha=\phi/\phi_0=\phi/(2\pi)$.
The model is periodic in $\alpha$ with a period $1$. 
$\vec u_z$ is the unit vector perpendicular to the plane. $\vec r_i$ are the 
positions of the vortices in the plane~: they are uncorrelated random  
variables (Poisson distribution).
The $\delta$ interactions (coupling to the magnetic field) are necessary 
in order to define properly the model 
\cite{JacPi90,McCOuv91,Ouv94,BerLoz94,ComMasOuv95}.

By comparison with scalar impurities discussed in the introduction, the
magnetic nature of the scatterers gives rise to rather different
properties\footnote{
  For scalar impurities, the presence of a magnetic field also strongly
  affects the spectral properties. The case of a Gaussian disorder projected
  in the lowest Landau level (LLL) of a strong magnetic field has been studied
  in ref.~\cite{Weg83}. Other disordered potentials have been considered
  later.  In particular it was shown in ref.~\cite{BreGroItz84} 
  that for a weak density of impurities, the spectrum can display power law
  singularities. A physical interpretation of such power law singularities has
  been discussed by Furtlehner in ref.~\cite{Fur97,Fur00}.  The relation
  between the model of scalar impurities projected in the LLL of a strong
  magnetic field and the model of magnetic vortices was discussed in
  refs.~\cite{DesFurOuv96,Fur97}.  Finally we mention a recent work on
  Lifshitz tails in the presence of magnetic field \cite{LesWar04}.
}~: it was shown in ref.~\cite{DesFurOuv95} that the spectrum is
reminiscent of a Landau spectrum with Landau levels broadened by disorder, in
the limit of vanishing magnetic flux $\alpha\to0$ (see below). This analysis
was performed using perturbative arguments supported by numerical
simulations.
The latter use a relation between the average DoS of
this particular model and some winding properties of the Brownian motion.

Let $Z(t)$ denotes the partition function for a given distribution of vortices
and $Z_0(t)$ the partition function without fluxes. The ratio of partition
functions can be written as a ratio of two path integrals
\be\label{pi1}
\frac{Z(t)}{Z_0(t)}=
\frac{
  \int \D \vec a 
  \int_{\vec r(0)=\vec a}^{\vec r(t)=\vec a}{\cal D}\vec r(\tau) 
  \:\EXP{\int_0^t (-\frac{1}{2} \dot{\vec r}^2 + \I\vec A\cdot\dot{\vec r}) 
    \D\tau }
     }
     {
  \int \D\vec a
  \int_{\vec r(0)=\vec a}^{\vec r(t)=\vec a}{\cal D}\vec r(\tau) 
  \:\EXP{-\int_0^t \frac{1}{2}\dot{\vec r}^2  \D\tau } 
     }
=
\left\langle 
  \EXP{ \I\oint_{\cal C}  \vec A \cdot \D\vec r}
\right\rangle_{\cal C}
\:,\ee
where $Z_0(t)={V}/({2\pi t})$ is the free partition function ($V$ is the
(infinite) area of the plane) and $\smean{\cdots}_{\cal C}$ stands for 
averaging over all closed Brownian curves of the plane.

In order to average $Z(t)$ over the Poissonian distribution of vortices, let
us consider, for a while, our problem on a square lattice with lattice spacing
$a$. Let $N_i$ be the number of vortices in square $i$. 
the magnetic flux through any closed random walk ${\cal C}$ on this lattice 
can be written as~:
\be\label{cl}
\oint_{\cal C}  \vec A \cdot \D \vec r = \sum_i 2 \pi \alpha N_i n_i 
\ee
where $n_i$ is the number of times the square $i$ has been wound around by 
${\cal C}$.

Averaging $Z(t)$ with the Poisson distribution ($\mu$ being the mean density of
vortices)
\be\label{pois}
 P(N_i) \; = \; \frac{ ( \mu a^2 )^{N_i} }{N_i!}\,\EXP{- \mu a^2 }  
\ee
gives
\be\label{zbar}
\mean{ Z(t) }_{ \{\vec r_i\} } 
=Z_0(t) 
\left\langle 
  \EXP{ \I\oint_{\cal C}  \vec A \cdot \D\vec r}
\right\rangle_{ {\cal C},\{\vec r_i\} }
=
Z_0(t) 
\left\langle   
\exp\left(     
 \mu \sum_n S_n(\EXP{2\I\pi\alpha n} - 1)
 \right) 
\right\rangle_{\cal C}
\ee
where $\smean{\cdots}_{\{\vec r_i\}}$ denotes averaging over the positions
$\{\vec r_i\}$ of the vortices.

\begin{figure}[!ht]
\begin{center}
\includegraphics[scale=.4,angle=0]{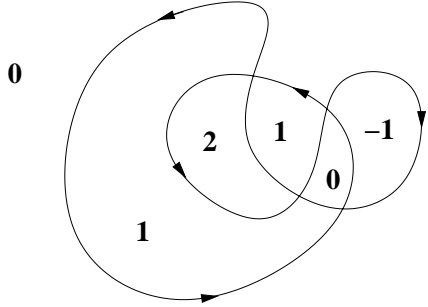}
\caption{
  \it A closed curve with its $n$-winding sectors. The label $n$
  of a sector is  the winding number of any point inside 
  this sector. \label{fig:jean1}}
\end{center}
\end{figure}

The quantity $S_n$ denotes the area of the locus of points around which the
curve ${\cal C}$ has wound $n$ times. This result was derived for a random
walk on a lattice but is also obviously valid off lattice as well, for 
Brownian curves. The winding sectors of a closed
curve are displayed in figure~\ref{fig:jean1}.

\noindent
The random variable $S_n$ scales like $t$.  In particular, we know
\cite{ComDesOuv90,GroFri03} its expectation $\smean{S_n}=\frac{t}{2\pi n^2}$.
Moreover, it has been shown in  \cite{Wer94} that the variable
$n^2S_n$ becomes more and more peaked when $n$ grows:
$
P(n^2 S_n = X) \leadto{n\to\infty} \delta(X-\frac{t}{2\pi })
$ (the average area of the $n=0$-sector has been recently studied in 
ref.~\cite{GarTru05}).

Thus, extracting   $t$, eq.~(\ref{zbar}) rewrites as
\be\label{zbar1}
\mean{ Z(t) }_{ \{\vec r_i\} }  = Z_0(t) 
\int\D S\D A\:P(S,A) \:\EXP{- \mu t ( S + \I A )}
\equiv 
Z_0(t) 
\left\langle \EXP{- \mu t ( S + \I A )}\right\rangle_{\cal C}
\:,\ee
where  $P(S,A)$ is the joint distribution of the 
rescaled ($t$ independent) variables $S$ and $A$ defined as
\bea
\label{s} 
S&=&\frac{2}{t} \sum_n S_n \sin^2 (\pi \alpha n)\\
\label{a}   
A&=&\frac{1}{t} \sum_n S_n \sin (2\pi \alpha n) 
\:.\eea
The averages of these two variables are given by~:
\bea
   \mean{S} &=& \pi\alpha (1 - \alpha) 
\\
   \mean{A} &=& 0
\:.\eea
With eq.~(\ref{zbar1}), we observe that $\frac1V\mean{Z(t)}_{\{\vec r_i\}}$
has the scaling form ${F(\mu t)}/{t}$. Thus, its inverse Laplace transform,
the average density of states per unit area, is a function of only $E/\mu$ and
$\alpha$. Moreover, it is easy to realize that $\mean{Z(t)}_{\{\vec r_i\}}$ is
even in $\alpha$ (each Brownian curve in $\{{\cal C}\}$ comes with its time
reversed) and periodic in $\alpha$ with period 1. Thus, one can restrict to
$0\leq\alpha\leq\frac{1}{2}$. Let us focus on the two limiting cases 
$\alpha\to0$ and $\alpha={1}/{2}$.

\vspace{0.25cm}

\begin{figure}[!h]
\begin{center}

\includegraphics[scale=0.4]{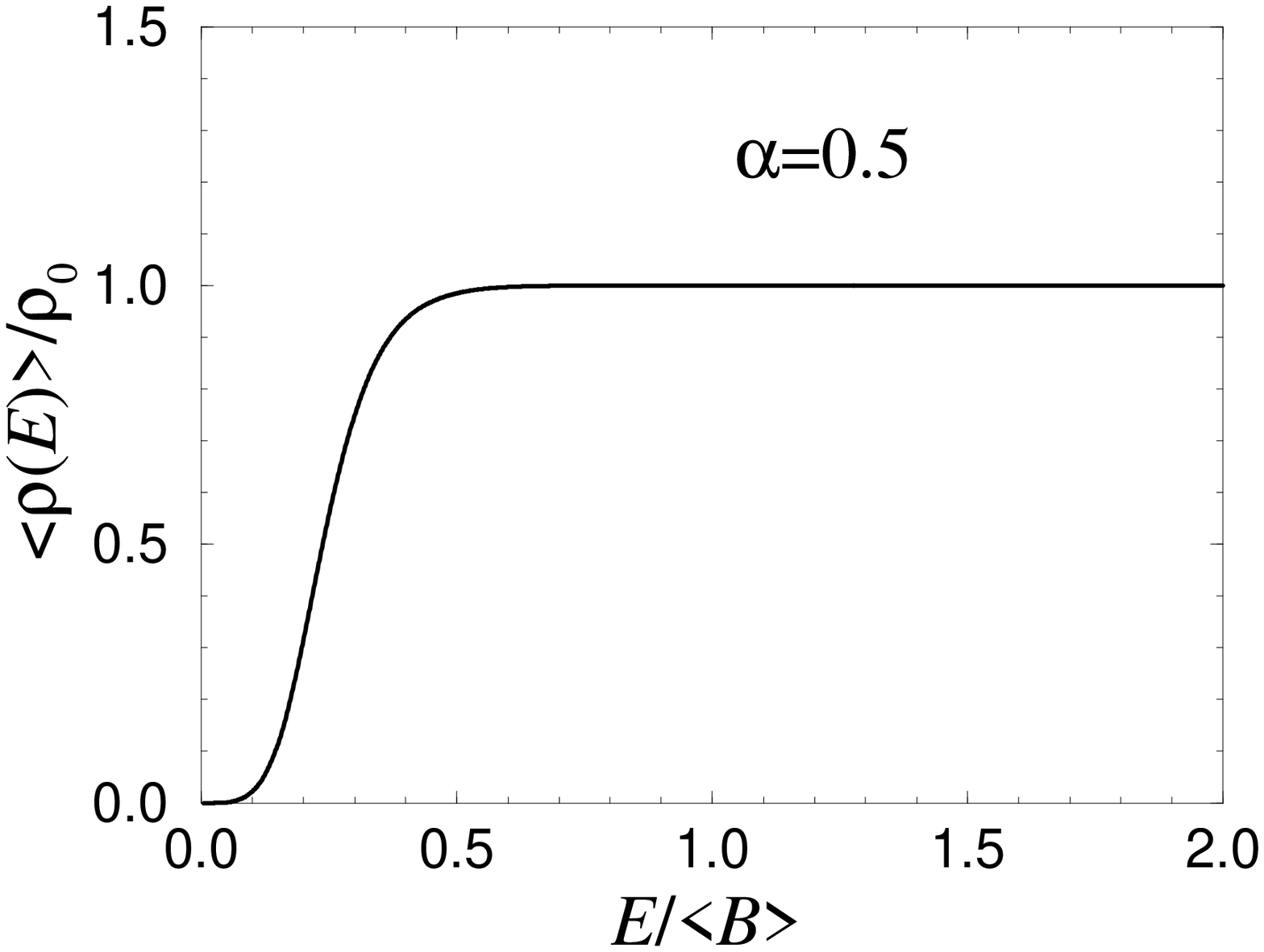}
\hspace{0.5cm}
\includegraphics[scale=0.4]{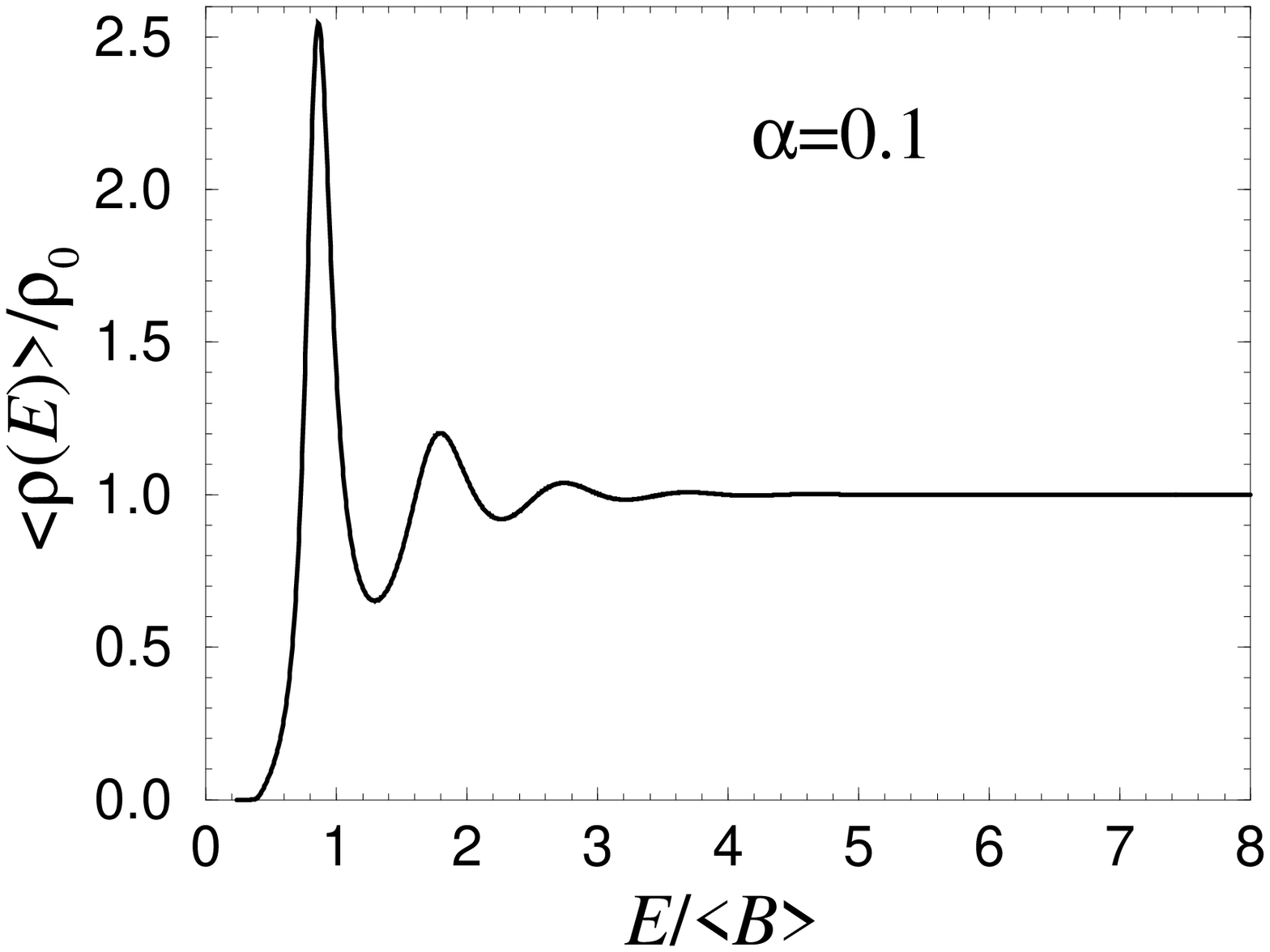}

\vspace{0.25cm}

\includegraphics[scale=0.4]{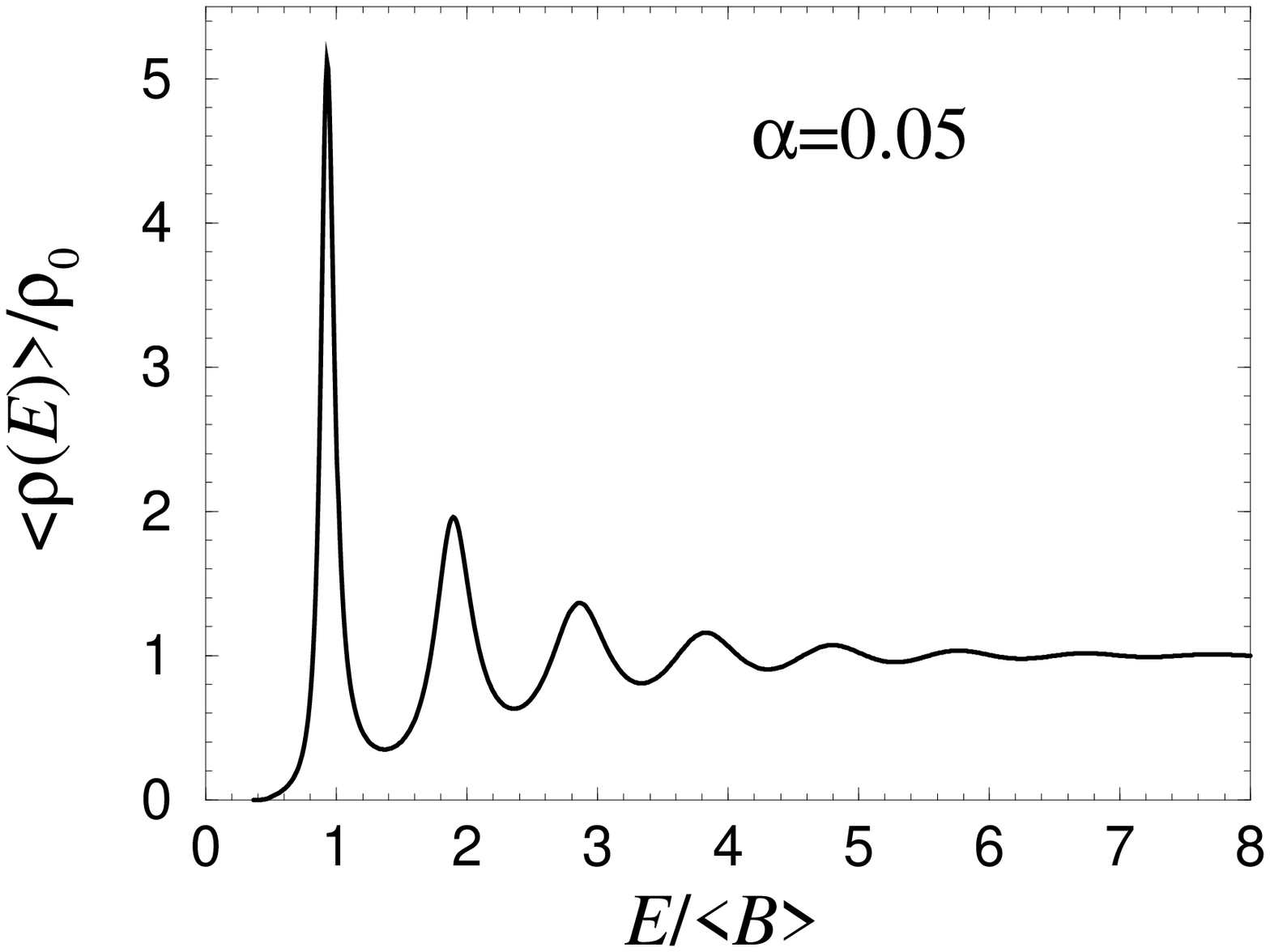}
\hspace{0.5cm}
\includegraphics[scale=0.4]{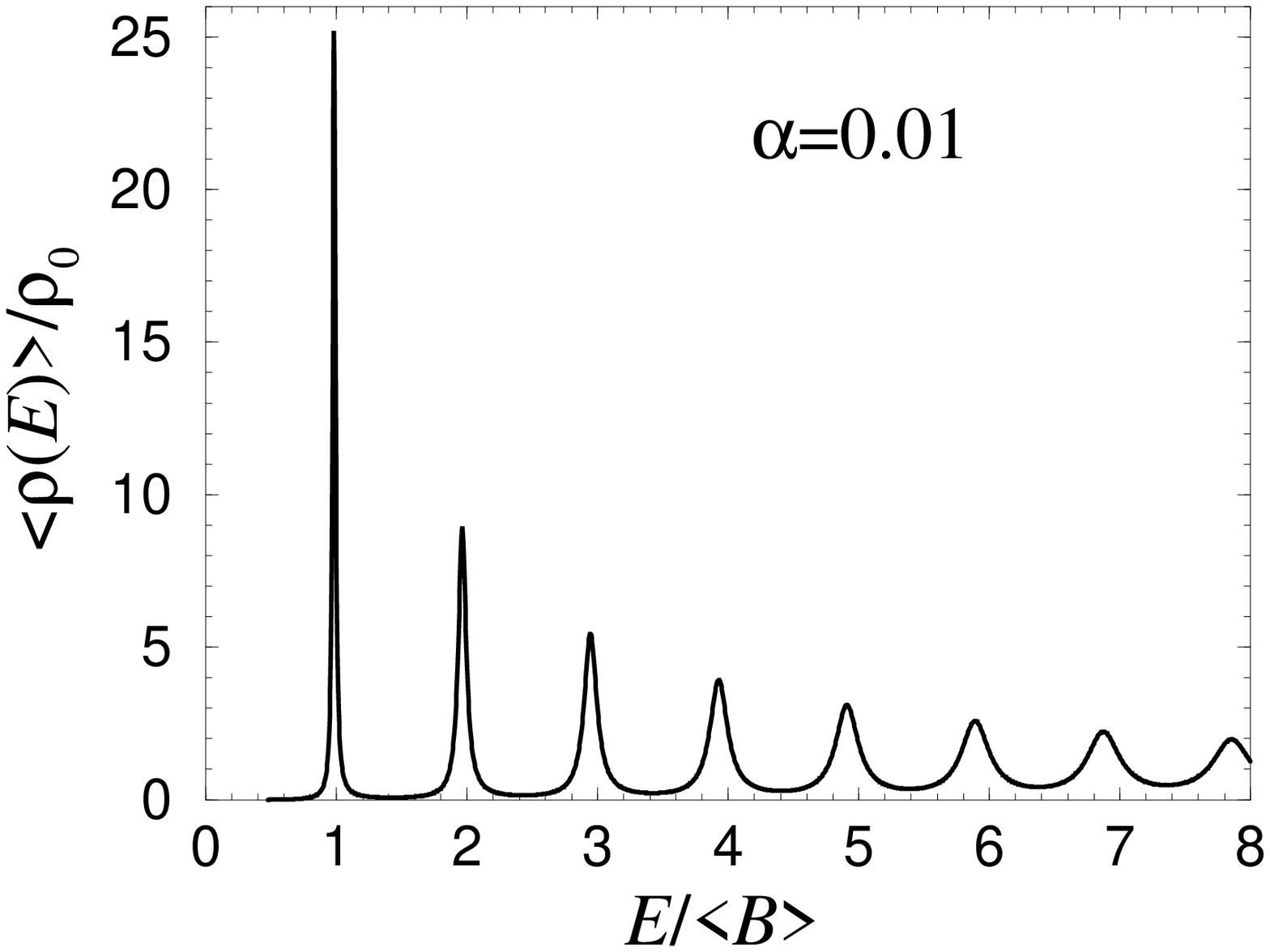}

\caption{\it Average density of states of the Hamiltonian (\ref{Him}) 
         for different values
         of the flux per tube $\phi=\alpha\phi_0$. 
         The average magnetic field reads $\mean{{\cal B}}=2\pi\mu\alpha$.
         We recall that the corresponding Landau spectrum is given by 
         $E_n=\mean{{\cal B}}(n+1)$, for $n\geq0$.
         From ref.~\protect\cite{DesFurOuv95}.\label{fig:dosimpmag}}
\end{center}
\end{figure}

\mathversion{bold}
\noindent$\bullet$ {\bf Limit $\alpha \to 0$~: Landau spectrum.}\\
\mathversion{normal}
When  $\alpha \to 0$, a careful analysis shows that
\be\label{atou0}
\mean{Z(t)}_{ \{\vec r_i\} } 
\simeq 
Z_{\smean{{\cal B}}} \: \EXP{-\frac12\smean{{\cal B}}t }
\ee
where 
$
Z_{\smean{{\cal B}}}=Z_0(t)
\frac{\smean{{\cal B}}t/2}{\sinh(\smean{{\cal B}}t/2)}
$
is the partition function of a charged particle in a uniform magnetic field
$\mean{{\cal B}}=2\pi\mu\alpha$ (we recall that the impurity $i$ carries a
magnetic field $2\pi\alpha\delta(\vec r-\vec r_i)$). The system of random
vortices is, thus, equivalent to the uniform average magnetic field, albeit 
with  an additional positive shift in the Landau spectrum~: the 
inverse Laplace transform of the partition function (\ref{atou0})
gives the Landau spectrum made of equally spaced infinitely degenerated
levels. This corresponds to the oscillating behaviour shown on
figure~\ref{fig:dosimpmag}.  The origin of the shift can be traced back to
the presence of the repulsive $\delta$ interactions that have been added to
the Hamiltonian to define properly the model.
   
\vspace{0.25cm}

\mathversion{bold}
\noindent$\bullet$ 
{\bf Half quantum flux vortices ($\alpha=1/2$) -- The spectral singularity at $E=0$.}\\
\mathversion{normal}
In this case the variable (\ref{a}) vanishes, $A\equiv0$, implying that
$
\mean{Z(t)}_{\{\vec r_i\}} = Z_0(t) 
\smean{ \exp \left( - \mu t  S  \right) }_{\cal C}
$ 
where now
\be\label{snodd}
 S= \frac{2}{t} \sum_{n\ {\rm odd}} S_n  
\:.\ee
Performing  
the inverse Laplace transform, we get the average density of states
  per unit area:
\be\label{dos1/2} 
\left\langle \rho (E)
 \right\rangle = \rho_0 (E)\ \int_0^{E/\mu} \D S\, P(S)
\ee
where $\rho_0(E)=\frac{1}{2\pi}$  is the free density of states per unit area
and  $P(S)$ is the probability distribution of $S$.

\begin{figure}[!ht]
\begin{center}
\includegraphics[scale=.5,angle=0]{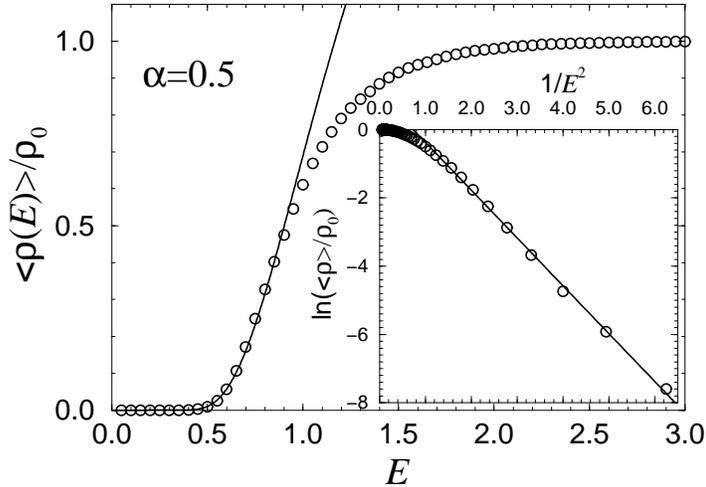}
\caption{\it 
 The  average density of states at $\alpha =0.5$ (the free density of states
 is constant $\rho_0=1/(2\pi)$). Circles are  simulation
 results. The full line is the fit discussed in the  text.
 \label{fig:jean2}}
\end{center}
\end{figure}

We may use this result to determine the nature of the singularity in the
average DoS. As displayed in figure~\ref{fig:jean2} (where we have taken
$\mu=1$), $\smean{\rho(E)}$ increases monotonically from $0$ to $\rho_0(E)$
with a depletion of states at the bottom of the spectrum. Circles are the 
result
of numerical simulations (on a 2D square lattice, we have generated 10000
closed random walks of 100000 steps each). The full line is a low-energy
fit of the quantity $\smean{\rho(E)}/\rho_0$ by the function
\begin{equation}
  \label{specsing}
  \frac{\rho_{\rm fit}(E)}{\rho_0} = a\, \EXP{-b(\mu/E)^2}
\end{equation}
with $a=2.8$ and $b=1.4$ (see figure~\ref{fig:jean2}).
It is worth stressing that this behaviour is quite different from the one 
expected for a disorder due to a scalar potential. The famous Lifshitz 
argument, applicable to the low energy DoS for a low concentration of 
{\it scalar} impurity, leads instead to\footnote{
  The Lifshitz argument applies to the DoS for the random Hamiltonian
  $H=-\Delta+\sum_i u(\vec r-\vec r_i)$ in dimension~$d$, 
  where $u(\vec r)$ is a sharply peaked scalar potential.
  For a low density of impurities $\mu$, the DoS behaves as 
  $\rho_{\rm Lif.}(E) \sim \exp({-c_d\,\mu/E^{d/2}})$ at low energy.
  \label{footnoteLif}
}
a behaviour 
\begin{equation}
\rho_{\rm Lif.}(E) \sim \EXP{-c_2\,\mu/E}\:,
\end{equation}
where $c_2$ is a constant. Our choice for this fit is motivated by a recent
numerical work \cite{Ric03} concerning the area $\cal{A}$ of the outer
boundary of planar random loops. In this work, the author suggests that the
limit distribution of $\cal{A}$ is the Airy distribution implying, for small
$\cal{A}$ values, a behaviour of the type 
$\exp(-\mbox{const.}/{\cal A}^2)/{\cal A}^2$. Remarking that 
${\cal A}=\sum_{n\ {\rm even}}S_n+\sum_{n\ {\rm odd}}S_n$, it is natural to
expect for the distribution of the random variable $S$, eq.~(\ref{snodd}), a
behaviour at small $S$ that is roughly given by $\exp(-\mbox{const.}/S^2)$.
Thus, we deduce the form $\rho_{\rm fit}(E)$ for the low-energy fit of
$\smean{\rho(E)}$.

\vspace{0.25cm}

\mathversion{bold}
\noindent$\bullet$ {\bf Transition between $\alpha=0.5$ and $\alpha\to0$.}\\
\mathversion{normal}
Finally, when $\alpha$ grows from $0$ to $0.5$, the oscillations in the
spectrum must disappear at some critical value $\alpha_c$ (see
figure~\ref{fig:dosimpmag}). Numerical simulations, specific heat
considerations and, also, diagrammatic expansions \cite{DesFurOuv95} give
$\alpha_c\sim0.3$.

\section{Conclusion}

A non experienced reader may have the feeling that the two topics which have
been covered in this review, namely one-dimensional disordered systems and
quantum graphs, are essentially disjoint. In fact there are many interesting
links between these two topics, both at a methodological and a conceptual
level. The use of metric graphs for modelling quantum phenomena observed in
disordered metals goes back to the pioneering work of Shapiro \cite{Sha82}
and Chalker \& Coddington \cite{ChaCod88}. These systems have been used to
study quantum localization and more recently as a model system for spectral
statistics (for a recent review, see ref.~\cite{Tan05}).  Another set of
similar questions is provided by the study of scattering properties of chaotic
graphs \cite{KotSmi03} and disordered systems \cite{FyoSom97} (see also
section~\ref{sec:wtd} and ref.~\cite{OssFyo05} for recent developments). We
have also seen that apart from its interest to study the spectral properties
of metric graphs, the spectral determinant also allows studying several
properties of networks of quasi-one-dimensional weakly disordered 
wires~\cite{Pas98,PasMon99,AkkComDesMonTex00}.

As a conclusion we would like to mention several open problems~:
\\
$\bullet$ A study of spectral statistics in the case of graphs with a random
Schr\"odinger operator [there is still a factorised structure but the matrix
$R$ is now given by eq.~(\ref{n41})].
\\
$\bullet$ A probabilistic understanding of the star graphs using the tools of
excursion theory developed by Barlow {\it et al}. \cite{BarPitYor89} in the
context of the Brownian spider. A first step would be to recover those
probabilistic results (the joint law of the occupation time inside the
branches) by a spectral approach. It is however not excluded that the
probabilistic approach could provide a key to a deeper understanding of those
quantum systems.  In the context of classical systems such probabilistic
approaches have been very useful, {\it e.g.} recent studies of the Stochastic
Loewner Equation have made enormous progress in understanding the statistical
physics of a class of two-dimensional systems. This calls for new
probabilistic techniques for quantum systems as well.
\\
$\bullet$ Several functionals of the Brownian motion (\ref{A},\ref{functring})
appear when studying the important question of dephasing due to
electron-electron interaction in networks of quasi-one-dimensional weakly
disordered wires. The fact that such simple functionals appear relies on the
translation invariance of the two particular problems studied (an infinite
wire \cite{AltAroKhm82} and an isolated ring \cite{TexMon05}). For a network
with arbitrary topology, the relevant functional of the Brownian bridge
$x(\tau)$ is given by $\int_0^t\D\tau\,W(x(\tau),x(t-\tau))$ where
$W(x,x')=\frac12[P_d(x,x)+P_d(x',x')]-P_d(x,x')$ with
$-\Delta{}P_d(x,x')=\delta(x-x')$.  It now involves a nonlocal functional in
time which is difficult to handle. Progress in this direction would allow 
clarifying the interplay between the electron-electron interaction and the
geometrical effect and help in analyzing recent experimental results
\cite{FerAngRowGueBouTexMonMai04,BauMalMonSamSchTex05}.
\\
$\bullet$ The question of extreme value spectral statistics was addressed in
the framework of random matrix theory \cite{TraWid93} and these studies have
found several applications in the context of out-of-equilibrium statistical
physics (see the review \cite{TraWid02}). However this question was first
addressed in the context of one-dimensional disordered systems
\cite{GreMolSud83}. The study of supersymmetric random Hamiltonian, for which
the bottom of the spectrum plays a special role, has emphasized the interest
in extreme value spectral statistics \cite{Tex00} (in particular it indicates
level correlations). It would be interesting to extend such studies to other
models and find some physical situations where these results would be
applicable.
\\
$\bullet$ A beautiful heuristic argument was provided by Lifshitz to explain
the spectral singularity for Hamiltonians with a weak concentration of
localized scalar impurities (see for example the book \cite{LifGrePas88}).
Other mechanisms should be invoked to explain the nature of the quantum states
responsible for the low energy power law behaviour in the case of a uniform
strong magnetic field with $\delta$-impurities \cite{Fur97,Fur00}.  For the
model of randomly distributed magnetic fluxes, the numerical simulations for
$\alpha=1/2$ have suggested the new type of singular behaviour
(\ref{specsing}). It would be interesting to provide heuristic arguments to
understand more deeply the origin of this singularity.


\section*{Acknowledgments}

The article gives an overview of several joint works involving~: \'Eric
Akkermans, Cyril Furtlehner, Satya Majumdar, Gilles Montambaux, C\'ecile
Monthus, St\'ephane Ouvry and Marc Yor. We thank them for fruitful
collaborations. Meanwhile we have also benefited from stimulating discussions
with Marc Bocquet, Eug\`ene Bogomolny, Oriol Bohigas, H\'el\`ene Bouchiat,
Markus B\"uttiker, David Dean, Richard Deblock, Meydi Ferrier, Sophie
Gu\'eron, Jean-Marc Luck, Gleb Oshanin, Leonid Pastur and Denis Ullmo.
We are pleased to acknowledge Satya Majumdar for his careful reading and 
valuable remarks.


\newpage


\end{document}